\documentstyle [10pt,eqsecnum,aps,amsfonts] {revtex}
\input epsf
\newcommand{\beq}{\begin{equation}}
\newcommand{\eeq}{\end{equation}}
\newcommand{\beqs}{\begin{eqnarray}}
\newcommand{\eeqs}{\end{eqnarray}}

\begin{document}
\draft

\baselineskip 5.0mm

\title{Ground State Degeneracy of Potts Antiferromagnets: Cases with 
Noncompact $W$ Boundaries Having Multiple Points at $1/q=0$}

\author{Robert Shrock\thanks{email: shrock@insti.physics.sunysb.edu}
\and Shan-Ho Tsai\thanks{current address: Department of Physics and Astronomy,
University of Georgia, Athens, GA 30602; email: tsai@hal.physast.uga.edu}}

\address{
Institute for Theoretical Physics  \\
State University of New York       \\
Stony Brook, N. Y. 11794-3840}

\maketitle

\vspace{10mm}

\begin{abstract}

We present exact calculations of the zero-temperature partition function, 
$Z(G,q,T=0)$, and ground-state degeneracy (per site), $W(\{G\},q)$, for the 
$q$-state Potts antiferromagnet on a number of families of graphs $\{G\}$ 
for which the boundary ${\cal B}$ of regions of analyticity of $W$ in the 
complex $q$ plane is noncompact and has the properties that (i) in the 
$z=1/q$ plane, the point $z=0$ is a multiple point on ${\cal B}$ and (ii) 
${\cal B}$ includes support for $Re(q) < 0$.  These families are generated by
the method of homeomorphic expansion.  Our results give further insight into
the conditions for the validity of large--$q$ series expansions for the reduced
function $W_{red.}=q^{-1}W$.  

\end{abstract}


\vspace{16mm}

\pagestyle{empty}
\newpage

\pagestyle{plain}
\pagenumbering{arabic}
\renewcommand{\thefootnote}{\arabic{footnote}}
\setcounter{footnote}{0}

\section{Introduction}

   This paper continues our study of nonzero ground state entropy, 
$S_0(\{G\},q) \ne 0$, i.e., ground state degeneracy (per site) 
$W(\{G\},q) > 1$, where 
$S_0 = k_B \ln W$, in $q$-state Potts antiferromagnets \cite{potts,wurev} 
on various lattices and, more generally, families of graphs $\{G\}$. 
There is an interesting 
connection with graph theory here, since the zero-temperature partition
function of the above-mentioned $q$-state Potts antiferromagnet on a graph
$G$ satisfies $Z(G,q,T=0)_{PAF}=P(G,q)$, where $P(G,q)$ is the chromatic
polynomial expressing the number of ways of coloring the vertices of the graph
$G$ with $q$ colors such that no two adjacent vertices have the same color 
\cite{birk}--\cite{biggsbook}.  Thus,
\beq
W([\lim_{n \to \infty} G \ ],q) = \lim_{n \to \infty} P(G,q)^{1/n}
\label{w}
\eeq
where $n=v(G)$ is the number of vertices of $G$. 
An example of a real substance exhibiting nonzero ground 
state entropy is ice \cite{lp,liebwu}.  Just as complex analysis
provides deeper insights into real analysis in mathematics, the generalization
from $q \in {\mathbb Z}_+$, to $q \in {\mathbb C}$ yields a deeper
understanding of the behavior of $W(\{G\},q)$ for physical (positive integral) 
$q$.  In general, $W(\{G\},q)$ is an analytic function in the $q$ plane 
except along a certain continuous locus of points, which we denote ${\cal B}$. 
In the limit as $n \to \infty$, the locus ${\cal B}$ forms by means of a 
coalescence of a subset of the zeros of $P(G,q)$ (called chromatic zeros of 
$G$).  \cite{early} 
In a series of papers we have calculated and analyzed $W(\{G\},q)$ for a 
variety of families of graphs \cite{p3afhc}--\cite{strip}, both for 
physical values of $q$ (via rigorous upper and lower bounds, large--$q$ series
calculations, and Monte Carlo measurements) and for the generalization to
complex values of $q$.  

A basic question that one can ask about the locus 
${\cal B}$ for (the $n \to \infty$ limit of a) family of graphs $\{G\}$ is
whether it extends only a finite distance from the origin of the $q$ plane or,
instead, extends an infinite distance away from this origin 
(i.e., passes through the point $z=0$ in the $z=1/q$ plane), 
and hence is noncompact.  The
importance of this question for the statistical mechanics of $q$-state Potts
antiferromagnets is that large--$q$ Taylor series provide a powerful means of
obtaining approximate values of the ground state degeneracy even for moderate
values of $q$ \cite{nagle}--\cite{kewser}, \cite{ww,wn}, 
but these exist if and only if the reduced function \footnote{
This reduced function is a natural object to define since 
an obvious upper bound on $P(G,q)$ describing the coloring of an $n$-vertex 
graph with $q$ colors is $P(G,q) \le q^n$, and hence $W(\{G\},q) \le q$; this
guarantees that $lim_{q \to \infty} W_{red.}(\{G\},q)$ is a finite quantity
(even if it is sometimes nonanalytic at this point).  
The large--$q$ series take their simplest form as series in the expansion
variable $y=1/(q-1)$ for the equivalent reduced function 
$\overline W(\{G\},q) = q^{-1}(1-q^{-1})^{-\Delta/2}W(\{G\},q)$, where $\Delta$
is the coordination number of the lattice.} 
$W_{red.}(\{G\},q) = q^{-1}W(\{G\},q)$ is analytic at $1/q=0$, which, in turn,
is true if and only if the nonanalytic boundary ${\cal B}$ does not pass
through $z=0$.  Large--$q$ series calculations of $W_{red.}$ have been 
derived for regular lattices \cite{nagle}, but we noted \cite{w} that for the
$r \to \infty$ limit of the graph $\overline K_2 + C_r$ the 
locus ${\cal B}$, which is noncompact \cite{read91}, no such large--$q$ series
exists. Here, $\overline K_p$ denotes the complement of the 
complete graph $K_p$ \footnote{
The complete graph $K_n$ on $n$ vertices is defined as the
graph where each of these vertices is connected to all of the other $n-1$
vertices by bonds.  The complement $\overline K_n$ of $K_n$ is the graph with 
$n$ vertices and no bonds.} 
It is clearly important to understand better the differences between the
behavior of Potts antiferromagnets on families of graphs that yield 
$W_{red.}(\{G\},q)$ functions analytic at
$z=0$ and those that do not, i.e. those that 
yield noncompact loci ${\cal B}$ passing
through $z=0$ \cite{wa}.  In the present paper, we carry out such a study
for families of graphs which are 
homeomorphic expansions of the following type: we start with families of the 
form $(K_p)_b + G_r$, 
where $G_r$ is a given graph family with $r$ vertices, $b$
signifies the removal of certain bonds in the $K_p$ subgraph, and 
$G+H$ is the ``join'' of the graphs $G$ and $H$ \footnote{
The ``join'' of two graphs $G$ and $H$ is obtained
by adding bonds connecting each of the vertices of $G$ to those of $H$.  This
was denoted $G \times H$ in Ref. \cite{wa}; here we use
the more common notation for this object in the mathematical literature, viz.,
$G+H$.} 
We then perform \underline homeomorphic \underline expansion of the bonds 
\underline connecting the $K_p$ and $G_r$ subgraphs.  This family is
categorized as being of type $HEC$.  An interesting feature of the families
analyzed here is that, in contrast to those that we have previously studied,
they yield loci ${\cal B}$ (which are boundaries of regions of analyticity of
$W$) that have support for $Re(q) < 0$ and for which the point $z=0$ is a
multiple point on the algebraic curve ${\cal B}$ in the technical terminology
of algebraic geometry \cite{alg}, i.e., a point where several branches of
this curve cross each other.  For
basic definitions on homeomorphic classes of graphs, see Ref. \cite{biggsbook};
some recent work on homeomorphism classes in a different direction than ours 
is Ref. \cite{rw}.  

   A general form for the chromatic polynomial of an $n$-vertex graph $G$ is
\beq
P(G,q) =  c_0(q) + \sum_{j=1}^{N_a} c_j(q)a_j(q)^{t_j n}
\label{pgsum}
\eeq
where $c_j(q)$ and $a_j(q)$ are certain functions of $q$. Here the $a_j(q)$ and
$c_{j \ne 0}(q)$ are independent of $n$, while $c_0(q)$ may contain
$n$-dependent terms, such as $(-1)^n$, but does not grow with $n$ like
$(const.)^n$.   A term $a_\ell(q)$ is defined as ``leading'' if it dominates
the $n \to \infty$ limit of $P(G,q)$; in particular, if $N_a \ge 2$, then it
satisfies $|a_\ell(q)| \ge 1$ and $|a_\ell(q)| > |a_j(q)|$ for $j \ne \ell$, so
that $|W|=|a_\ell|^{t_j}$.  The locus ${\cal B}$ occurs where 
there is a nonanalytic change in $W$ as the leading terms $a_\ell$
in eq. (\ref{pgsum}) changes. 

   Since for the families of graphs studied here, the boundary ${\cal B}$
is noncompact in the $q$ plane, it is often more convenient to describe the
boundary in the complex $z$ or $y$ planes, where, as above, $z=1/q$, and 
\beq
y \equiv \frac{1}{q-1} = \frac{z}{1-z}
\label{y}
\eeq
The variable $y$ is commonly used in large--$q$ series expansions.  For the
families considered here ${\cal B}$ is
actually compact in the $y$ and $z$ planes.  We define polar coordinates as
\beq
z = \zeta e^{i\theta}
\label{zpolar}
\eeq
and
\beq
y = \rho e^{i\beta}
\label{ypolar}
\eeq
and the function 
\beq
D_k(q) = \frac{P(C_k,q)}{q(q-1)} = a^{k-2}\sum_{j=0}^{k-2}(-a)^{-j} =
\sum_{s=0}^{k-2}(-1)^s {{k-1}\choose {s}} q^{k-2-s}
\label{dk}
\eeq
where $a=q-1=1/y$ and $P(C_k,q)$ is the chromatic polynomial for the circuit 
(cyclic) graph $C_k$ with $k$ vertices,
\beq
P(C_k,q) = a^k + (-1)^ka
\label{pck}
\eeq
We shall also use a standard notation from combinatorics, 
\beq
q^{(p)} = p! {q \choose q-p} = \prod_{s=0}^{p-1}(q-s)
\label{ff}
\eeq

The organization of the paper is as follows. In section 2 we construct several
general multiparameter homeomorphic classes of families of graphs that, in a
certain limit, yield noncompact boundaries ${\cal B}(q)$.  In sections 3-5 
we give exact calculations of the respective boundaries ${\cal B}$ for three
such families.  These exhibit features going beyond those that we found in 
our earlier work in three main ways: (i) the point $z=0$ can be a multiple 
point; (ii) ${\cal B}$ includes support for $Re(q) < 0$, and (iii) it is no
longer true in general that $Re(q) =0 \ \Leftrightarrow q=0$ for $q \in {\cal
B}$.  A general discussion of our findings is given in section 6, and our
conclusions are presented in the final section.

\section{Homeomorphic Classes of $HEC$ Type}

\begin{figure}
\centering
\leavevmode
\epsfxsize=3.5in
\epsffile{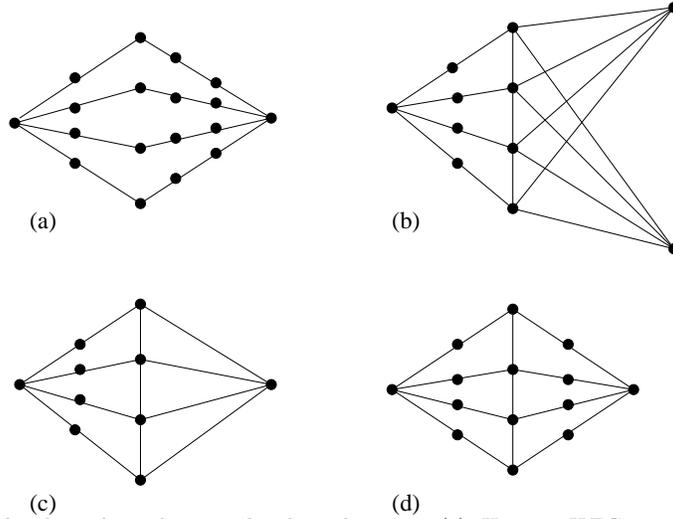}
\caption{\footnotesize{Illustrations of families of graphs considered in the
text: (a) $H_{k,r} = HEC_{k_1-2,k_2-2}(\overline K_2 + \overline K_r)$, 
with $r=4$ and $k_1=3$, $k_2=4$, whence $k=k_1+k_2-1=6$; 
(b) $\Theta_{k_1,r} = HEC_{k_1-2,0,0}(\overline K_3 + T_r)$ with 
$k_1=3$ and $r=4$; 
(c) $O_{k_1,r} = HEC_{k_1-2,0}(\overline K_2 + T_r)$ with $k_1=3$ and 
$r=4$; (d) $U_{k,r} = HEC_{k_1-2,k_2-2}(\overline K_2 + T_r)$ with 
$k_1=k_2=k=3$ and $r=4$.  Here, $K_p$ is the complete graph with $p$ 
vertices and $T_r$ is the tree graph with $r$ vertices.}}
\label{hhegraph1}
\end{figure}

We can construct a large variety of families of graphs with noncompact 
boundaries ${\cal B}(q)$ of regions where the respective $W$ functions are
analytic by performing homeomorphic expansions of the basic family 
constructed and analyzed before \cite{wa}, 
\beq
(K_p)_b + G_r \ , 
\label{kpbgr}
\eeq
where, as before, $b$ denotes the removal of $b$ bonds connecting a vertex 
$v$ in the $K_p$ subgraph to the other vertices of $K_p$.  Since each vertex
$v \in K_p$ has degree $\Delta = p-1$, $b$ is bounded above according to 
$b \le p-1$.  We have shown earlier that (i) the locus 
${\cal B}$ for $\lim_{r \to \infty}[(K_p)_b + G_r]$ is noncompact in
the $q$ plane, passing through $z=1/q=0$ \cite{wa} and (ii) the analogous locus
for the $r \to \infty$ limit of homeomorphic expansions of this family has the
same noncompactness property.  To construct specific homeomorphic
expansions, select a vertex $v_1$ in 
the $K_p$ subgraph and perform homeomorphic expansion on each of the bonds 
connecting this vertex with the vertices in $G_r$ by inserting $k_1-2$ 
additional degree-2 vertices on each of these bonds, where $k_1 \ge 3$.  Then
continue this process with a second vertex $v_2 \in K_p$, inserting 
$k_2-2$ additional degree-2 vertices on each of bonds connecting $v_2$ with the
vertices of $G_r$, and so forth for the other vertices in $K_p$ (including the
vertex $v$ from which the $b$ bonds were removed, if there are any bonds left
linking with this vertex). We denote the resultant homeomorphic expansion as 
\beq
HEC_{k_1-2,k_2-2,...,k_p-2}[(K_p)_{b(v)} + G_r]
\label{heckpbgr}
\eeq
The labelling is chosen so that, counting the two vertices on the original
bond connecting to $v_j \in K_p$ together with the $k_j-2$ inserted
vertices, there is a total of $k_j$ vertices in all along what was originally
this single bond.  Some illustrative examples of this and other types of
families to be discussed are shown in Fig. \ref{hhegraph1}. 

Another starting family with a noncompact ${\cal B}$ in the $r \to \infty$
limit is 
\beq
(K_p)_{\{b\};\{v\}} + G_r
\label{kppgr}
\eeq
where the subscript $\{b\}$ refers to 
the removal of multiple bonds from a set of non-adjacent vertices $\{v\}$ 
of the $K_p$ subgraph. Performing the $HEC$-type homeomorphic 
expansion as above leads to the family
\beq
HEC_{k_1-2,k_2-2,...,k_p-2}[(K_p)_{\{b\};\{v\}} + G_r] 
\label{heckpbbgr}
\eeq

A special case of eq.(\ref{heckpbbgr}) applies if in the starting family one
removes all of the bonds connecting vertices in the $K_p$ subgraph to each 
other, so that this starting family is 
\beq
\overline K_p + G_r 
\label{epgr}
\eeq
The $HEC$-type homeomorphic expansion of this family is thus 
\beq
HEC_{k_1-2,k_2-2,...,k_p-2}(\overline K_p + G_r)
\label{hecepgr}
\eeq
In this case one clearly obtains the same homeomorphically expanded family 
if one permutes the choices of $k_j$:
\beq
HEC_{k_1-2,k_2-2,...,k_p-2}(\overline K_p + G_r) = 
HEC_{\pi(k_1)-2,\pi(k_2)-2,...,\pi(k_p)-2}(\overline K_p + G_r)
\label{hecepgrsymmetry}
\eeq
where $\pi$ is an element of the permutation group on $p$ objects, $S_p$. 
The number of vertices in the homeomorphic classes of families 
(\ref{heckpbgr}), (\ref{heckpbbgr}), and (\ref{hecepgr}) is the same, namely 
\beqs 
v \Bigl ( HEC_{k_1-2,k_2-2,...,k_p-2}[(K_p)_b + G_r] \Bigr ) & = & 
v \Bigl ( HEC_{k_1-2,k_2-2,...,k_p-2}[(K_p)_{\{b\};\{v\}} + G_r ] 
\Big ) 
\cr & = & 
v \Bigl ( HEC_{k_1-2,k_2-2,...,k_p-2}[\overline K_p + G_r ] \Big ) 
\cr
& = & p + r(1-2p + \sum_{j=1}^p k_j) 
\label{vheckpbgr}
\eeqs
Note that for $p=2$, since one can only remove $b=1$ bond in the $K_2$
subgraph, thereby obtaining $(K_2)_{b=1} = \overline K_2$, it follows also 
that 
\beq
HEC_{k_1-2,k_2-2}[(K_2)_{b=1} + G_r] = 
HEC_{k_1-2,k_2-2}[\overline K_2 + G_r]
\label{hecp2b}
\eeq

For the case $G_r = \overline K_r$, the chromatic number is 
\beq
\chi \Bigl ( HEC_{k_1-2,k_2-2,...,k_p-2}[\overline K_p + \overline K_r] 
\Bigr ) = 2
\label{chiheceper}
\eeq
For $G_r = T_r$, for the cases $r \ge 2$ of interest here, 
\beq
\chi \Bigl ( HEC_{k_1-2,k_2-2,...,k_p-2}[\overline K_p + T_r]
\Bigr ) = 3
\label{chiheceptr}
\eeq
so that for $r \ge 2$, 
\beq
P(HEC_{k_1-2,k_2-2,...,k_p-2}[\overline K_p + T_r],q=2)=0
\label{peptrq20}
\eeq
For the $r=1$ case, this family degenerates to a tree graph,
\beq
HEC_{k_1-2,k_2-2,...,k_p-2}[\overline K_p + T_1] = 
T_{v_t}  \quad {\rm where} \quad v_t = 1+\sum_{j=1}^p (k_j-1)
\label{heceptr1tree}
\eeq
with chromatic polynomial $P(T_{v_t},q)=q(q-1)^{v_t-1}$ and chromatic 
number $\chi=2$.

A general result for the chromatic polynomial in the case $r=2$, $G_r=T_r$ is
\beq
P \Bigl ( HEC_{k_1-2,k_2-2,...,k_p-2}[\overline K_p + T_2],q \Bigr ) = 
q(q-1)\prod_{j=1}^p D_{2k_j-1}
\label{hectr2}
\eeq
Since the respective indices $2k_j-1$ of each of the $D_{2k_j-1}$ in eq. 
(\ref{hectr2}) are odd and since 
\beq
D_{k \ odd}=(q-2) \times pol(q)
\label{dkkoddfactor}
\eeq
where $pol(q)$ is a polynomial in $q$ of degree $k-3$, it follows that 
\beq
P \Bigl ( HEC_{k_1-2,k_2-2,...,k_p-2}[\overline K_p + T_2],q \Bigr ) =
q(q-1)(q-2)^p \times pol'(q)
\label{hectr2factorq2}
\eeq
where $pol'(q)$ is a polynomial of degree $2\sum_{j=1}^p(k_j-2)$. 

Since the number of vertices 
$n=v$ is a linear function of $p$, $r$, and $k_1,...,k_p$, there are
several ways of producing the limit $n \to \infty$ ($L$ denotes limit): 
\beq
L_p: \ p \to \infty \quad {\rm with} \quad r \ , k_1 \ , ..., \ k_p  \quad
{\rm fixed}
\label{pinf}
\eeq
\beq
L_r : \ r \to \infty \quad {\rm with} \quad p \ , k_1 \ , ..., \ k_p  \quad 
{\rm fixed}
\label{rinf}
\eeq
\beq
L_{k_j}: \ k_j \to \infty \quad {\rm with} \quad p \ , k_1 \ , .., k_{j-1}, 
k_{j+1}, ...k_p \quad {\rm fixed}
\label{kjinf}
\eeq
and, for the case where all of the $k_j$'s are the same, i.e., 
$k_j = k_{cb} \ \forall j$ (where the subscript $cb$ denotes ``connecting
bond''), 
\beq
L_k: \ k_{cb} \to \infty \quad {\rm with} \quad p, \ r \quad {\rm fixed} 
\label{kinf}
\eeq
As discussed before \cite{wa}, the limit $L_p$ is not of much interest from the
viewpoint of either statistical mechanics or graph theory, since for any 
given graph $G_r$ and for any given (finite) value of
$q \in {\mathbb Z}_+$, as $p$ becomes sufficiently large, one will not be able
to color the graph $(K_p)_b + G_r$ or homeomorphic expansions
thereof, and the chromatic polynomial will vanish.  The limits $L_{k_j}$ and 
$L_k$ are also not of primary interest here since they generically yield
compact boundaries ${\cal B}$, as will be illustrated below.  
We shall therefore concentrate on the limit $L_r$.  The lowest choice
of $p$ that yields a noncompact boundary ${\cal B}(q)$ is $p=2$.

For general $p$ and $G_r$, we remark on two special types of $HEC$ 
homeomorphic expansions.  A minimal case is one in which the above
homeomorphic expansion is carried out only on the bonds connecting a single
vertex, taken to be $v_1$ with no loss of generality, in $\overline K_p$ to 
vertices in $G_r$, so $k_2=k_3=...=k_p=2$: 
\beq
HEC_{k_1-2,0,,...,0}(\overline K_p + G_r) 
\label{hecepgrv1}
\eeq
A particularly symmetric case is the one in which the homeomorphic expansions 
are the same on all of the connecting bonds: 
\beq
HEC_{k_1-2,k_2-2,...,k_p-2}(\overline K_p + G_r) \quad {\rm with}
\quad k_1 = k_2 = ... = k_p \equiv k_{cb} 
\label{hecepgrksame}
\eeq
In this case, the right-hand side of eq. (\ref{vheckpbgr}) reduces to 
$v=p + r + pr(k_{cb}-2)$. 
We proceed to give detailed analyses of some special families of homeomorphic 
expansions.

\section{Class of Families $HEC_{\lowercase{k}_1-2,\lowercase{k}_2-2}
(\overline K_{\lowercase{2}} + \overline K_{\lowercase{r}})$}

The family $HEC_{k_1-2,k_2-2}(\overline K_2 + \overline K_r)$ has the special
property that the graphs of this family only depend on the sum of 
$k_1$ and $k_2$, not on each of these parameters individually: 
\beq
HEC_{k_1-2,k_2-2}(\overline K_2 + \overline K_r) = 
HEC_{k_1'-2,k_2'-2}(\overline K_2 + \overline K_r)  \quad \Leftrightarrow 
\quad k_1 + k_2 = k_1' + k_2'
\label{e2ersym}
\eeq
To incorporate this symmetry, we define
\beq
H_{k,r} = HEC_{k-3}(\overline K_2 + \overline K_r) = 
HEC_{k_1-2,k_2-2}(\overline K_2 + \overline K_r) \quad {\rm where} 
\quad k=k_1+k_2-1
\label{hkr}
\eeq
As will be evident in our explicit results to be given below, the fact that 
this family has a noncompact $W$ boundary ${\cal B}(q)$ in the $L_r$ limit 
can be understood to follow from its construction 
as a homeomorphic expansion of the starting family 
$(K_2)_{b=1} + G_r = \overline K_2 + G_r$ which (as was shown previously 
\cite{wa}) has 
a noncompact ${\cal B}$ (here, $G_r = \overline K_r$).  The 
noncompactness of the boundary ${\cal B}$ obviously implies that in the $L_r$
limit, the chromatic zeros have unbounded magnitudes (in the $q$ plane), since
${\cal B}$ arises via the coalescence of these chromatic zeros in this limit 
\cite{wa} \footnote{
For a family of graphs $G_{k,r}$, rather than analyzing the
continuous locus ${\cal B}$ resulting from the $L_r$ limit in eq. 
(\ref{rinf}) or the $L_k$ in eq. (\ref{kinf}),
one may formulate a different problem: let $\{q_0\}_{k,r}$ denote the set of 
chromatic zeros of $G_{k,r}$, and consider the union of this set, summed 
over both $r$ and $k$, denoted $\{q_0\}_{ \forall k,r} = 
\sum_{r=1}^\infty \sum_{k=2}^\infty \{q_0\}_{k,r}$.  This problem has been
considered for the family $G_{k,r}=H_{k,r}$ by A. Sokal, who finds (private 
communication) that the image of this union is dense in the 
vicinity of the origin, $y=z=0$.} 
We note that the case $k_1=k_2=2$, i.e., $k=3$ is a 
special case of the graphs with
noncompact ${\cal B}(q)$  that we have constructed and studied before 
\cite{wa}, so we concentrate on the homeomorphic expansions $k \ge 3$
here. The number of vertices is given by the $p=2$ special case of eq. 
(\ref{vheckpbgr}) with (\ref{hkr}), viz., $v(H_{k,r}) = (k-2)r+2$. 
For arbitrary $k \ge 2$ and $r \ge 1$, $H_{k,r}$ is bipartite, i.e., 
\beq
\chi(H_{k,r})=2
\label{chihkr}
\eeq
For $r \ge 2$, the girth is $g(H_{k,r}) = 2k-2$. 
Indeed, all (non-self-intersecting) closed paths are of this length.  
For $r=1$ and $r=2$, the family degenerates according to
\beq
H_{k,1}=T_k
\label{treedegen}
\eeq
and 
\beq
H_{k,2} = C_{2k-2}
\label{circuitdegen}
\eeq
where, as defined above, $T_n$ and $C_n$ are, respectively, the tree and
circuit graphs with $n$ vertices.  Hence, we shall take $r \ge 3$.  

By use of the deletion-contraction theorem \footnote{
We recall the statement of the addition-contraction theorem
\cite{rtrev}-\cite{biggsbook}: let $G$ be a graph, and let $v$ and $v'$ be two
non-adjacent vertices in $G$.  Form (i) the graph $G_{add.}$ in which one
adds a bond connecting $v$ and $v'$, and (ii) the graph $G_{contr.}$ in which
one identifies $v$ and $v'$. Then the chromatic polynomial for coloring $G$
with $q$ colors, $P(G,q)$, satisfies $P(G,q) = P(G_{add.},q) +
P(G_{contr.},q)$. The reverse process of bond deletion, leading to the same 
equation, is known as the deletion-contraction theorem.}
we obtain the recursion relation 
\beq
P(H_{k,r},q) = D_kP(H_{k,r-1},q) + (-1)^{k-1}q[(q-1)D_{k-1}]^{r-1}
\label{rec}
\eeq
which we solve to get the chromatic polynomial 
\beq
P(H_{k,r},q) = q(q-1)\biggl [ D_{2k-2}(D_k)^{r-2} -(q-1)(D_{k-1})^2
\Bigl [(D_k)^{r-2} - \bigl \{ (q-1)D_{k-1} \bigr \}^{r-2} \Bigr ] 
\biggr ]
\label{phkr}
\eeq
This has the form of eq. (\ref{pgsum}) with $N_a=2$, 
\beq
a_1 = D_k
\label{a1hkr}
\eeq
and 
\beq
a_2 = (q-1)D_{k-1}
\label{a2hkr}
\eeq
In the limit $L_r$ of eq. (\ref{rinf}), the nonanalytic boundary locus 
${\cal B}$ is determined as the solution of the degeneracy of magnitudes of the
leading terms 
\beq
|a_1|=|a_2|
\label{degeneqhkr}
\eeq
To determine ${\cal B}$, we multiply eq. (\ref{degeneqhkr}) by 
$|q(q-1)|$ to obtain $|P(C_k,q)|=|(q-1)P(C_{k-1},q)|$ (the resultant spurious 
solutions thereby introduced at $q=0,1$, i.e., $y=-1, \infty$, are understood
to be ignored in the following discussion).  Dividing by $|a|^k$ yields 
\beq
|1+(-1)^k y^{k-1}| = |1+(-1)^{k-1}y^{k-2}|
\label{degeneqy}
\eeq
Since $y=0$ is a solution of eq. (\ref{degeneqy}), ${\cal B}$ is noncompact 
in the $q$ plane, passing through $z=1/q=0$, as noted above 
(equivalently, eq. (\ref{degeneqhkr}) satisfies the condition given
in the theorem of section IV of Ref. \cite{wa} 
that guarantees that the solution locus
${\cal B}$ is noncompact in the $q$ plane).  
(In contrast, [for fixed $k$] ${\cal B}$ is compact in the $z$ or $y$ plane.) 

Furthermore, we find that (i) \ 
$2(k-2)$ curves on ${\cal B}$, consisting of $(k-2)$ 
complex-conjugate (c.c.) pairs, intersect at $z=y=0$; (ii) these 
curves approach the point $z=y=0$ at the angles 
\beq
\phi_j = \beta_j = \pm \frac{(2j+1)\pi}{2(k-2)} \ , \qquad j = 0,...,k-3
\label{angles}
\eeq
Thus, if and only if $k$ is odd, these angles include $\pm \pi/2$, i.e., a
branch of ${\cal B}$ crosses the point $z=y=0$ vertically. 
To prove these results, we write (\ref{degeneqy}) in terms of polar coordinates
using eq. (\ref{ypolar}): 
\beq
\rho^{k-2}\biggl [ \rho^{k-2}(\rho^2-1) + 2(-1)^k \Bigl \{ 
\rho \cos[(k-1)\beta] + \cos[(k-2)\beta] \Bigr \} \biggr ] =0
\label{polareq}
\eeq
As $\rho \to 0$, so that $z \to y$ (whence $\beta \to \phi$), the solution is 
given by $\cos[(k-2)\beta]=0$, which yields the results (i) and (ii). 
In the terminology
of algebraic geometry \cite{alg}, the point $z=y=0$ is a singular (multiple) 
point on the algebraic curve ${\cal B}$ of index $k-2$.

As a further result, we find that 
(iii) if $k$ is odd, ${\cal B}$ crosses the real $q$ axis once, at a
value $q_c$ that increases monotonically from 3/2 for $k=3$, with 
$\lim_{k \to \infty}q_c = 2$; (iv) 
if $k$ is even, ${\cal B}$ never crosses the real $q$ axis, so $q_c$ is not
defined. To show these results, we first 
set $\beta=0$ in eq. (\ref{polareq}), which becomes 
\beq
(\rho+1)^2\Bigl [ \rho^{k-2} + 2(-1)^k \sum_{s=0}^{k-3}(-\rho)^s \Bigr ] = 0
\label{eqcross}
\eeq
For odd $k$, this has a single real, positive solution for $\rho$, which
decreases monotonically from $\rho=2$ for $k=3$, approaching $\rho=1$ as $k \to
\infty$.  For even $k$, eq. (\ref{eqcross}) has no (real) solution for $\rho$.
For $\beta=\pi$, eq. (\ref{polareq}) yields
\beq
(\rho-1)^2\Bigl [ \rho^{k-2} + 2\sum_{s=0}^{k-3}\rho^s \Bigr ] = 0
\label{eqcrossbneg}
\eeq
Aside from the root $\rho=1$, i.e., $y=-1$, which is spurious as noted above,
eq. (\ref{eqcrossbneg}) has no real roots.  The result for $q$ follows 
directly. 

\pagebreak

\begin{figure}
\vspace{-4cm}
\centering
\leavevmode
\epsfxsize=3.5in
\begin{center}
\leavevmode
\epsffile{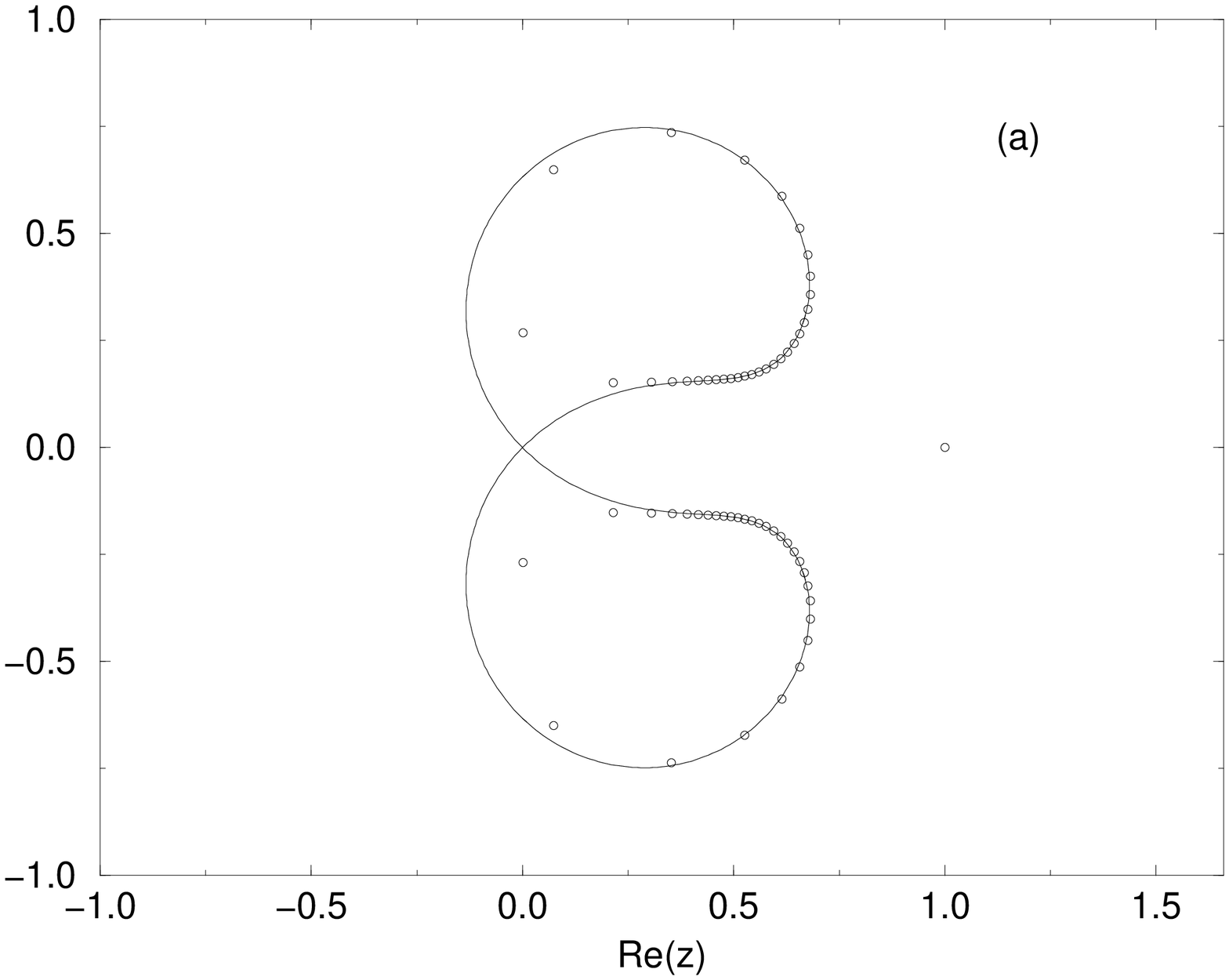}
\end{center}
\vspace{-4cm}
\begin{center}
\leavevmode
\epsfxsize=3.5in
\epsffile{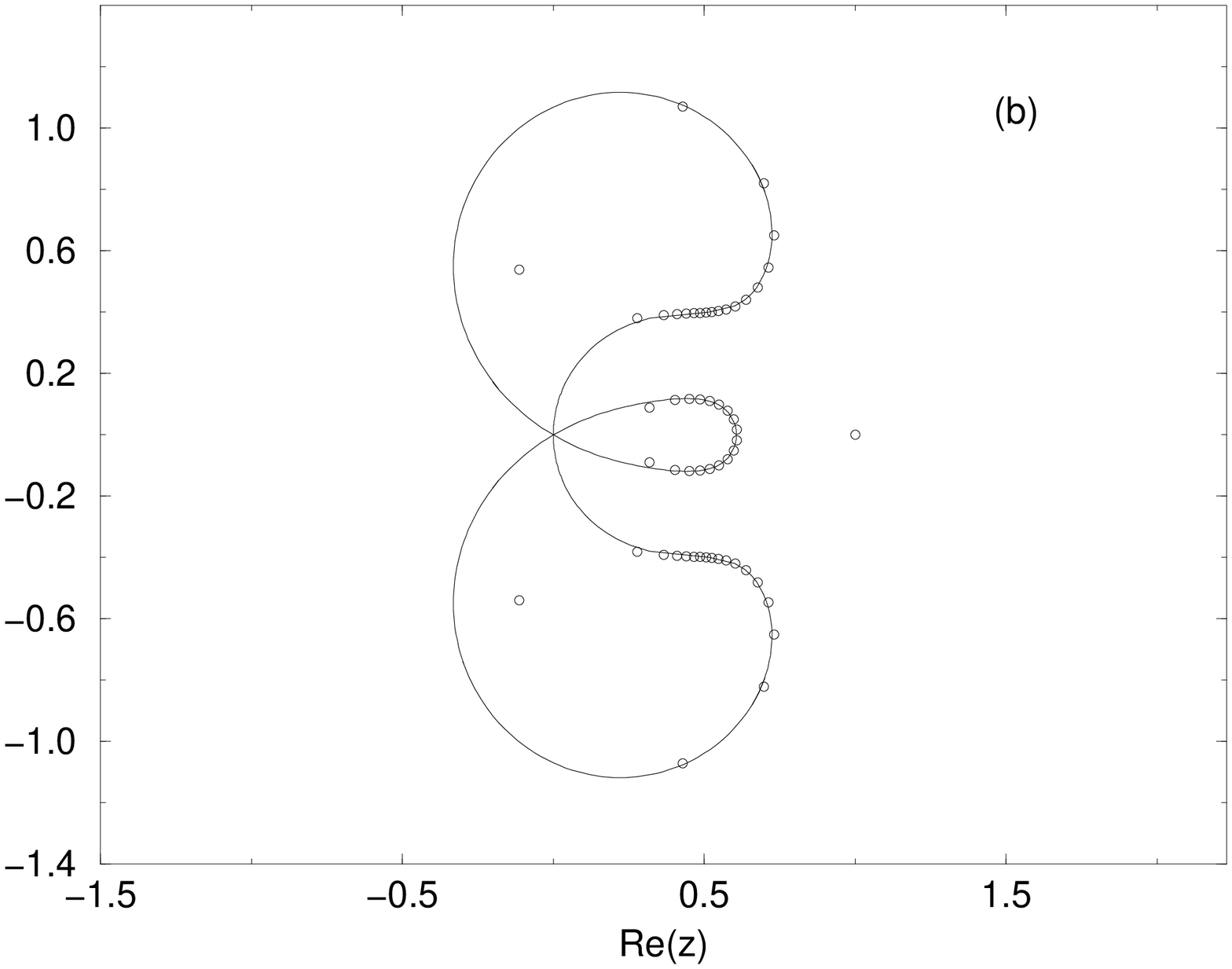}
\end{center}
\vspace{-4cm}
\begin{center}
\leavevmode
\epsfxsize=3.5in
\epsffile{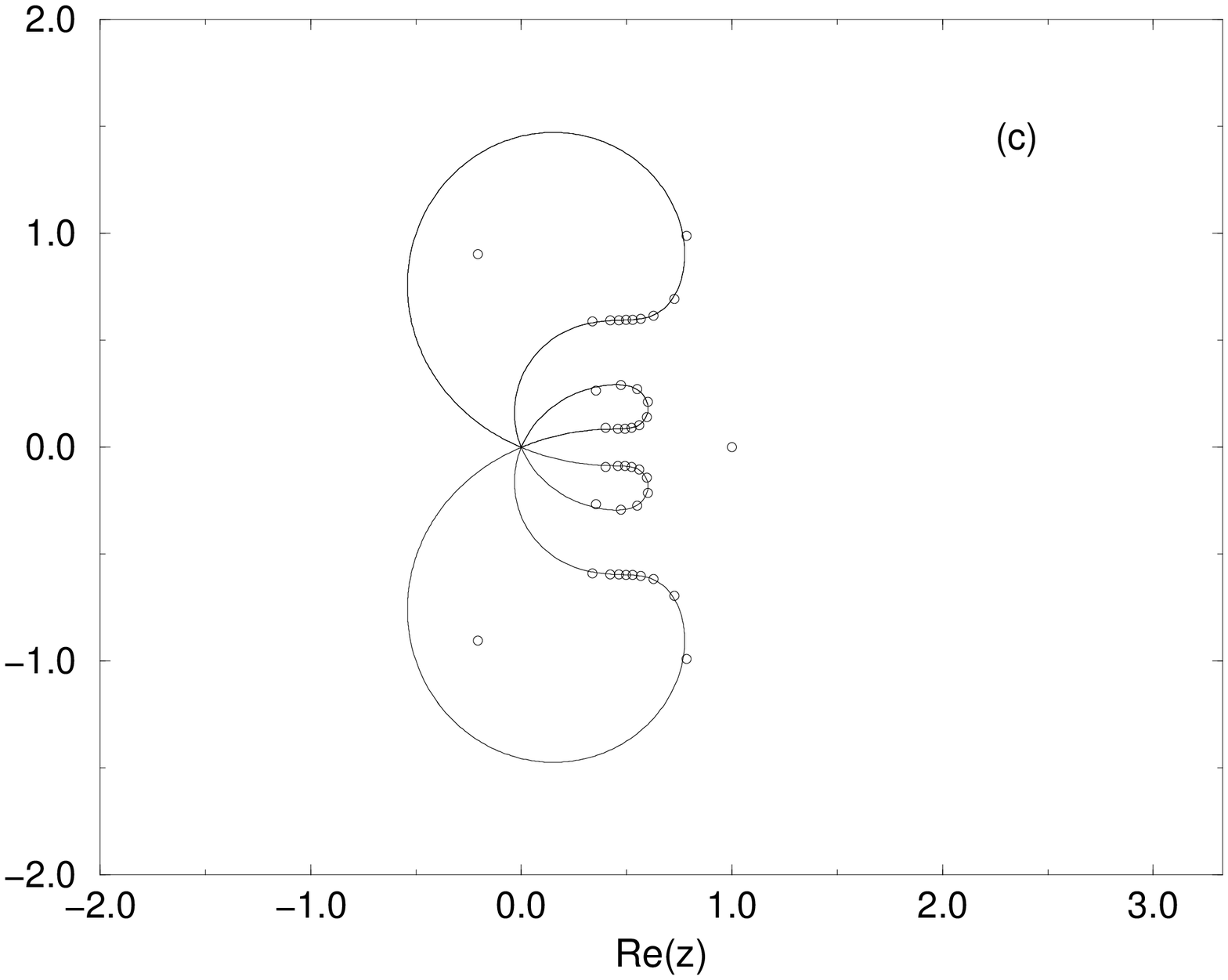}
\end{center}
\vspace{-2cm}
\caption{\footnotesize{Boundary ${\cal B}$ in the $z=1/q$ plane for 
$\lim_{r \to \infty}H_{k,r}$ with $k=$ (a) 4 (b) 5 (c) 6.  
Chromatic zeros for $H_{k,r}$ with $(k,r)=$ (a) (4,30) (b) (5,18) (c) (6,10) 
are shown for comparison.}}
\label{hamboundaryk4z}
\end{figure}

\pagebreak

\begin{figure}
\vspace{-4cm}
\centering
\leavevmode
\epsfxsize=3.5in
\begin{center}
\leavevmode
\epsffile{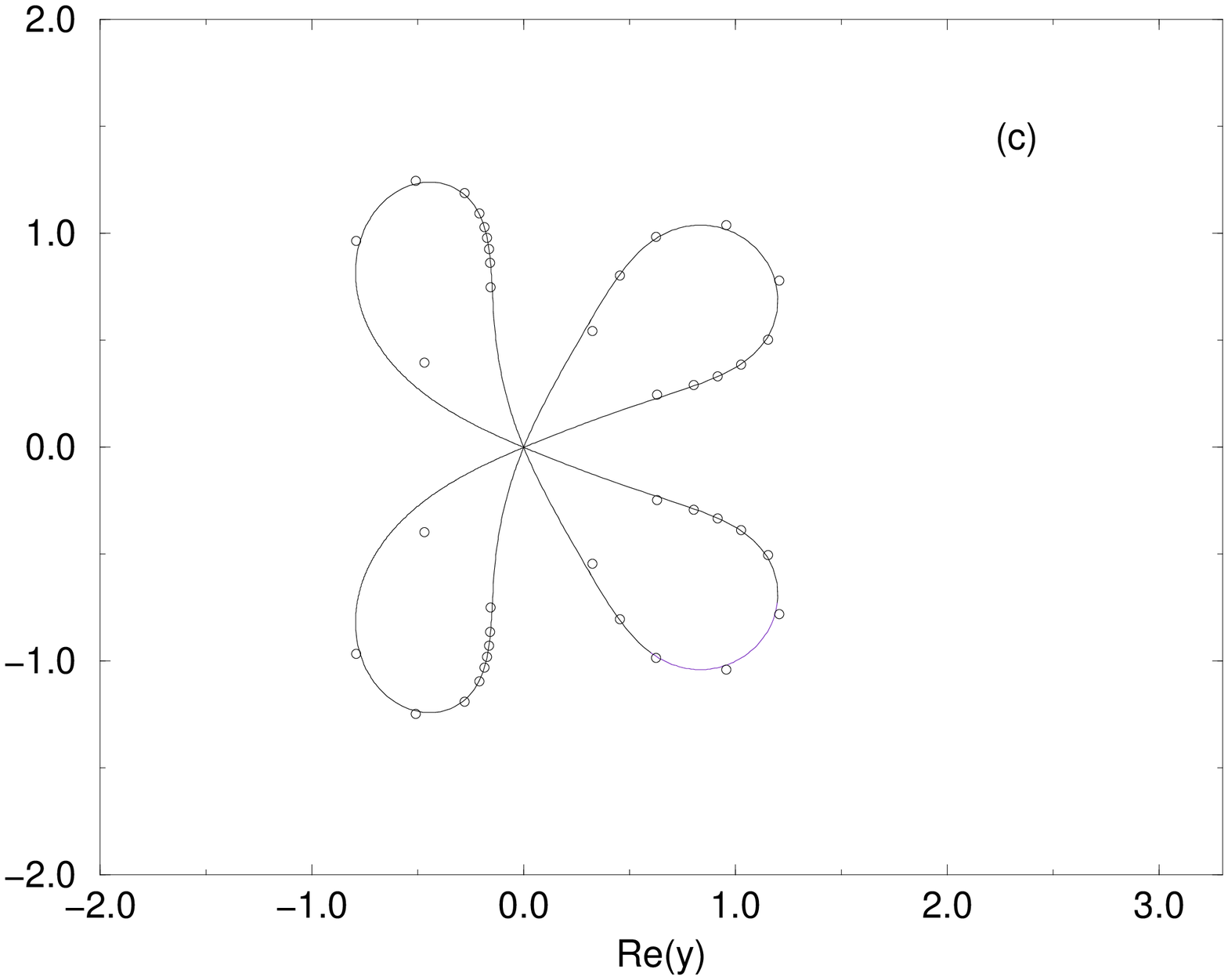}
\end{center}
\vspace{-2cm}
\caption{\footnotesize{Boundary ${\cal B}$ in the $y=1/(q-1)$ plane for 
$\lim_{r \to \infty}H_{k,r}$ with $k=6$.  Chromatic zeros 
for $H_{k,r}$ with $r=10$ are shown for comparison.}}
\label{hamboundaryk4y}
\end{figure}

The curves comprising ${\cal B}$ divide the $z$ (equivalently, $y$ or $q$) 
plane into $k-1$ regions.  For even $k=2\ell$ and odd $k=2\ell+1$, these 
include $(\ell-1)$ complex conjugate pairs of regions.  In the $q$ plane, 
${\cal B}$ consists of
$k-2$ disjoint curves (a line for $k=3$); each curve extends inward from
complex infinity, turns around and heads back out to complex infinity along a 
different direction.  For $k=3$, ${\cal B}$ is a circle in both the $y$ and 
$z$ planes, given by the equation $|y-1|=1$, i.e., 
$|z-\frac{1}{3}|=\frac{1}{3}$.   For $k=4,5,6$ we show our calculations of 
${\cal B}$ in the $z$ planes in Fig. \ref{hamboundaryk4z}.  The corresponding
plots in the $y$ plane look more like (slightly distorted) cloverleafs, 
reflecting the simplicity of the degeneracy equation (\ref{degeneqy}) in the 
$y$ variable; as an example, we show the case $k=6$ in Fig. 
\ref{hamboundaryk4y}.  (Additional figures in the $y$ plane are given in Ref. 
\cite{thesis}).  As is clear from Fig. 
\ref{hamboundaryk4z}, for $k \ge 4$, ${\cal B}$ has support for 
$Re(q) < 0$ (equivalent to $Re(z) < 0$).  
Note that $Re(z) = 0$ does not, in general, imply that $z=0$,
since for $k \ge 4$, ${\cal B}$ intersects the imaginary $z$ axis away from the
origin, e.g., at $z=\pm z_i$, where 
\beq
z_i= i\sqrt{2/5} = 0.632456i \quad {\rm for} \quad k=4
\label{zik4}
\eeq
\beq
z_i = i\sqrt{8/7} = 1.06904i \quad {\rm for} \quad k=5
\label{zik5}
\eeq
\beq
z_i = (i/3)[10 \mp \sqrt{82}]^{1/2} = 0.323971i, \ 1.45508i \quad {\rm for}
\quad k=6
\label{zik6}
\eeq
In the $y$ plane, the intercepts
$y=\pm i\rho_i$ of ${\cal B}$ with the imaginary $y$ axis (aside 
from $y=0$) are given, for even $k=2\ell$, by 
\beq
\rho_i^{2(\ell-1)}(\rho_i^2-1)-2(-1)^\ell=0
\label{yintercepteq}
\eeq
and, for odd $k=2\ell+1$, by eq. (\ref{yintercepteq}) multiplied by $\rho$.
Solving these equations for general $k$, we find that (with $k > 3$) 
(i) for $k=$ (0 or 1) mod 4, $\rho_i$ is nonzero, decreasing monotonically 
from $\sqrt{2}$ for $k=4,5$ toward 1 as $k \to \infty$; (ii) for 
$k=$ (2 or 3) mod 4, there is no nonzero intercept $\rho_i$. 
Although ${\cal B}$ extends to $Re(z) < 0$ (equivalently, $Re(q) < 0$), 
it never includes support for negative real $z$ or $q$.  

In Fig. \ref{hamboundaryk4z} we also show illustrative
chromatic zeros for (a) $k=4$, $r=30$ ($\Rightarrow \ n=62$), 
(b) $k=5$, $r=18$ ($\Rightarrow \ n=56$), and 
(c) $k=6$, $r=10$ ($\Rightarrow \ n=42$).  Aside from the
always-present zeros at $q=0,1$ ($z=\infty, 1$), the chromatic zeros 
generally lie close 
to the asymptotic curves comprising ${\cal B}$ onto which they coalesce as $k
\to \infty$.  For a given set of (finite) $k$ and $r$, the moduli of the 
zeros are bounded, and hence they avoid the points $z=y=0$, approaching these
only as $r \to \infty$ for a given $k$. 

 As before \cite{w,wa}, we let $R_1$ denote the region including the 
positive real $q$ axis for $q > q_c$.  For $k$ even, this includes the entire 
real $q$ axis. For odd $k$, we denote the region containing the rest of the 
real $q$ axis as $R_{2,k \ odd}$; from our statement above, it follows that
there is no $R_2$ phase for $k$ even.  For $q \in R_1$, we calculate 
\beq
W([\lim_{r \to \infty}H_{k,r}],q; q \in R_1) = 
\cases{ [a_2(q)]^{1/(k-2)} & if $k$ is odd \cr 
 [a_1(q)]^{1/(k-2)} & if $k$ is even \cr }
\label{wr1}
\eeq
As discussed in Ref. \cite{w}, in regions other than $R_1$, 
there is no canonical choice of phase in taking the $1/n$'th root in eq. 
(\ref{w}), so that one can only determine $|W(\{G\},q)|$.  We find
\beq
|W([\lim_{r \to \infty}H_{k,r}],q)| = |a_1(q)|^{1/(k-2)} \ , \quad {\rm for}
\quad k \ \ {\rm odd \ \ and} \quad q \in R_{2, k \ odd}
\label{koddwr2}
\eeq
\beqs
|W([\lim_{r \to \infty}H_{k,r}],q)| & = & |a_2(q)|^{1/(k-2)}  \ , 
\quad {\rm for} \quad k \ {\rm odd \ \ and} \quad q \not \in R_1, 
R_{2, k \ odd} \cr
& & \qquad \qquad \qquad 
{\rm and \ \ for} \quad k \ {\rm even \ \ and} \quad q \not\in R_1
\label{krj}
\eeqs
On ${\cal B}$, $|W|$ is continuous but nonanalytic.

\section{Class of Families $HEC_{\lowercase{k}_1-2,0,0}
(\overline K_{\lowercase{3}} + \overline K_{\lowercase{r}})$}

When $p \ge 3$ in the homeomorphic class of families (\ref{heckpbgr}) or 
(\ref{hecepgr}) with $G_r=\overline K_r$, the graphs and their chromatic 
polynomials depend on the individual values of the $k_j$, not just on the 
sum, as in eq. (\ref{e2ersym}).  A simple homeomorphic class of families 
of this type is the special case of eq. (\ref{hecepgrv1}) for $p=3$ and 
$G_r=\overline K_r$, namely 
\beq
\Theta_{k,r} = HEC_{k_1-2,0,0}(\overline K_3 + \overline K_r) \ , \quad 
k_1 \equiv k
\label{thetakr}
\eeq
We have $v(\Theta_{k,r})=r(k-1)+3$ and 
\beq
\chi(\Theta_{k,r})=2
\label{chitheta}
\eeq
For our present purposes, it will suffice to study the family $\Theta_{3,r}$. 
We calculate the chromatic polynomial 
\beqs
P(\Theta_{3,r},q) &=& (q-2)^{2r-4}P(\Theta_{3,2},q)-q(q-1)(D_4)^2 
\Bigl [ (q-2)^{2r-4}-(D_4)^{r-2} \Bigr ]  \cr\cr
& & -q(q-1)^2(q-2)^r \Bigl [ (q-2)^{r-2}-(q-1)^{r-2} \Bigr ] \cr\cr
& & -q(q-1)(q-2)(q^2-4q+5)^2\Bigl [ (q-2)^{2r-4}-(q^2-4q+5)^{r-2} \Bigr ] 
\label{pthetakr}
\eeqs
where
\beq
P(\Theta_{3,2},q)=q(q-1)\Bigl [ (q-1)D_6-(q-2)D_5 \Bigr ]
\eeq
This has the form of eq. (\ref{pgsum}) with $N_a=4$ and 
\beq
a_1 = D_4 = q^2-3q+3
\label{a1hece3er}
\eeq
\beq
a_2 = q^2-4q+5
\label{a2hece3er}
\eeq
\beq
a_3=(q-1)(q-2)
\label{a3hece3er}
\eeq
and
\beq
a_4 = (q-2)^2 
\label{a4hece3er}
\eeq

\begin{figure}
\centering
\leavevmode
\epsfxsize=3.5in
\begin{center}
\leavevmode
\epsffile{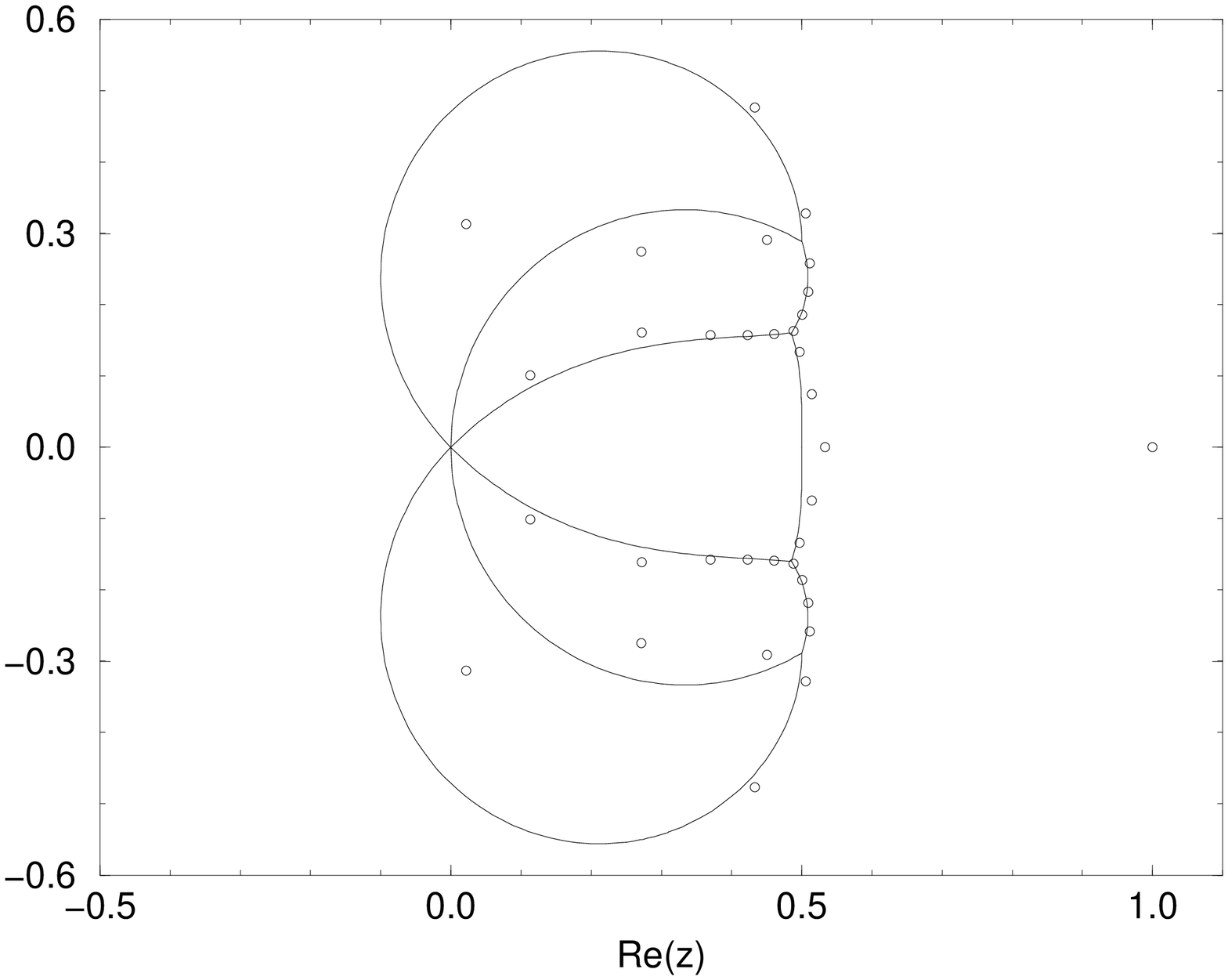}
\end{center}
\vspace{-2cm}
\caption{\footnotesize{Boundary ${\cal B}$ in the $z=1/q$ plane for 
$\lim_{r \to \infty}\Theta_{3,r}$.  Chromatic zeros 
for $\Theta_{3,r}$ with $r=16$ are shown for comparison.}}
\label{h1e3er}
\end{figure}
The nonanalytic boundary ${\cal B}$ for the limit as $r \to \infty$ is shown in
Fig. \ref{h1e3er}.  As with the $H_{k,r}$ family discussed above, and the other
families to be discussed below, one sees the
important feature that ${\cal B}$ includes support for $Re(z) < 0$ or
equivalently $Re(q) < 0$.  The boundary 
divides the $z$ (or equivalently $y$ or $q$) plane into
six different regions: (i) $R_1$, including the interval $0 < z < 1/2$ on the 
positive real $z$ axis; (ii) $R_2$, including the union of
the intervals $z > 1/2$ and $z < 0$; (iii) a complex-conjugate pair of regions
$R_3$ and $R_3^*$ lying just above and below $R_1$ in the $z$ plane; and 
(iv) a second pair of complex-conjugate regions $R_4$ and $R_4^*$ located such
that $R_4$ lies above and adjacent to $R_3$ (hence between $R_3$ and $R_2$) and
$R_4^*$ lies below $R_3^*$.  In the regions $R_1$ and $R_2$, $a_1$ and $a_2$
are dominant, respectively, while $a_j$ is dominant in regions $R_j$ and
$R_j^*$ for $j=3$ and $j=4$, respectively.  We have 
\beq
W([\lim_{r \to \infty}\Theta_{k,r}],q) = (a_1)^{1/2} \quad {\rm for} 
\quad q \in R_1
\label{wthetar1}
\eeq
\beq
|W([\lim_{r \to \infty}\Theta_{k,r}],q)| = |a_2|^{1/2} \quad {\rm for}
\quad q \in R_2
\label{wthetar2}
\eeq
and
\beq
|W([\lim_{r \to \infty}\Theta_{k,r}],q)| = |a_j|^{1/2} \quad {\rm for}
\quad q \in R_j, R_j^* \ , j=3,4 
\label{wthetar34}
\eeq

The region boundaries between regions $R_i$ and $R_j$ are the
solutions of the respective degeneracy equation $|a_i|=|a_j|$ where $a_i$ and
$a_j$ are leading terms.  Dividing these equations by $|q|^2$ and re-expressing
them in terms of $z$ to get the corresponding degeneracy equations in the $z$
plane, one has $|\tilde a_i|=|\tilde a_j|$ with $\tilde a_1 = 1-3z+3z^2$, 
$\tilde a_2 = 1-4z+5z^2$, $\tilde a_3 =(1-z)(1-2z)$, and $\tilde a_4 = 
(1-2z)^2$.  Clearly, each $(i,j)$ pair of the degeneracy equations 
$|\tilde a_i|=|\tilde a_j|$ in the $z$ plane has a solution at $z=0$, which
shows that, in accordance with the condition of Ref. \cite{wa}, ${\cal B}$ is
noncompact in the $q$ plane, passing through the point $z=0$.  Thus, as is
evident in Fig. \ref{h1e3er}, all of the regions are contiguous at $z=0$, 
where six curves (i.e., three branches) of ${\cal B}$ meet in a multiple 
point.  For small $\zeta = |z|$, the degeneracy equation for the part of ${\cal
B}$ separating $R_1$ and $R_3$ has the form $2\zeta^2(1-2\cos^2\theta) +
O(\zeta^3) = 0$, so that this boundary crosses the point $z=0$ at the angles
given by $\cos\theta=1/\sqrt{2}$, i.e., 
\beq
\theta_{R_1,R_3}, \ \theta_{R_1, R_3^*} = \pm \frac{\pi}{4}
\label{thetathetar1r3}
\eeq
(The other 
solution, $\cos\theta = -1/\sqrt{2}$ is not relevant because in this region,
$a_1$ and $a_3$ are not leading terms.)  Similarly, for small $\zeta$, the
degeneracy equation for the part of ${\cal B}$ separating the regions $R_3$ and
$R_4$ is $\zeta(3\zeta-2\cos\theta)=0$ so that as $\zeta \to 0$, this boundary
crosses the point $z=0$ at the angles given by $\cos\theta=0$, i.e., 
\beq
\theta_{R_3,R_4}, \ \theta_{R_3^*,R_4^*} = \pm \frac{\pi}{2}
\label{thetathetar3r4}
\eeq
The same type of reasoning applied to the $R_2,R_4$
and $R_2,R_4^*$ boundaries shows that they cross $z=0$ at the angles 
\beq
\theta_{R_2,R_4}, \ \theta_{R_2, R_4^*} = \pm \frac{3\pi}{4}
\label{thetathetar2r4}
\eeq
Concerning the boundary separating regions $R_1$ and $R_2$, we observe
that the relevant degeneracy equation 
\beq
{\cal B}(R_1 - R_2):  \ \ |1-3z+3z^2|=|1-4z+5z^2|
\label{bhece3erregion12}
\eeq
has the solution $z=1/2$ as its only solution other than $z=0$ where $a_1$ and
$a_2$ are leading terms.  Hence, $q_c=2$ for this family.  Concerning the two
complex-conjugate (c.c.) 
pairs of multiple points, we note that the multiple point $z_{2,3,4}$ 
where regions $R_2$, $R_3$, and $R_4$ are contiguous, and its c.c., are 
\beq
z_{2,3,4},z_{2,3,4}^* = 3^{-1/2}e^{\pm i\pi/6} = \frac{1}{2} \pm
 \frac{i}{2\sqrt{3}}
\label{mpr2r3r4}
\eeq
Hence this c.c. pair lies on the unit circle $|y|=1$ or equivalently 
$|q-1|=1$ in the respective $y$ and $q$ planes: 
\beq
y_{2,3,4}, \ y_{2,3,4}^* = e^{\pm i\pi/3}
\label{y234}
\eeq
\beq
q_{2,3,4}, \ q_{2,3,4}^* = 3^{1/2}e^{\pm i\pi/6}
\label{q234}
\eeq
A corresponding analysis can be given for the multiple point $z_{1,2,3}$ 
where regions $R_1$, $R_2$, and $R_3$ are
contiguous. 

\section{Class of Families $HEC_{\lowercase{k}_1-2,\lowercase{k}_2-2}
(\overline K_{\lowercase{2}} + T_{\lowercase{r}})$}

Illustrative graphs of the family 
\beq
HEC_{k_1-2,k_2-2}(\overline K_2 + T_r)
\label{heck2tr}
\eeq
with $(k_1,k_2)=(3,2)$ and $(k_1,k_2)=(3,3)$ are shown in
Fig. \ref{hhegraph1}(c,d), respectively. 
 From the general equation (\ref{vheckpbgr}) with $G_r=T_r$ and $p=2$, we have 
\beq
v \Bigl ( HEC_{k_1-2,k_2-2}(\overline K_2 + T_r) \Bigr ) = 2+r(k_1+k_2-3)
\label{vhece2tr}
\eeq
The special case $k_1=k_2=2$ is the family 
$\overline K_2 + T_r = (K_2)_{b=1} + T_r$ which we analyzed in detail earlier
\cite{wa} so we concentrate on the homeomorphic expansions of this starting
family here, i.e., $k_1 \ge 3$ and/or $k_2 \ge 3$.  In order to calculate the
chromatic polynomial for the graphs in this homeomorphic class of families, we
shall utilize a generating function method as we did in Ref. 
\cite{strip} for infinitely long, finite-width strip graphs of various 
lattices.  The generating function $\Gamma$ is a function of a
symbolic variable $x$ and yields the chromatic polynomials 
$P\Bigl (HEC_{k_1-2,k_2-2}(\overline K_2 + T_r),q \Bigr )$ 
via the Taylor series expansion around $x=0$ according to 
\beq
\Gamma \Bigl (HEC_{k_1-2,k_2-2}(\overline K_2 + T_r),q,x \Bigr ) = 
\sum_{r=1}^{\infty} P \Bigl (HEC_{k_1-2,k_2-2}(\overline K_2 + T_r),q
\Bigr ) x^r
\label{gammahectr}
\eeq
The summation starts at $r=1$ since this is the
minimum value of $r$ in this family. 
As in our earlier work \cite{strip}, we find that $\Gamma$ 
is a rational function of $x$ and, separately, of $q$, of the form
\beq
\Gamma \Bigl ( HEC_{k_1-2,k_2-2}(\overline K_2 + T_r),q,x \Bigr ) = 
\frac{{\cal N}\Bigl ( HEC_{k_1-2,k_2-2}(\overline K_2 + T_r),q,x \Bigr )}
{{\cal D}\Bigl ( HEC_{k_1-2,k_2-2}(\overline K_2 + T_r),q,x \Bigr )}
\label{gammagenhectr}
\eeq
with
\beq
{\cal N}\Bigl ( HEC_{k_1-2,k_2-2}(\overline K_2 + T_r),q,x \Bigr ) = 
x\sum_{j=0}^{j_{max}} A_j(q) x^j
\label{nhectr}
\eeq
(so that $deg_x({\cal N})=j_{max}+1$; the prefactor of $x$ reflects the 
fact that the minimum value of $r$ is 1) and
\beq
{\cal D}\Bigl ( HEC_{k_1-2,k_2-2}(\overline K_2 + T_r),q,x \Bigr ) = 
1 + \sum_{j'=1}^{j'_{max}} b_{j'}(q) x^{j'}
\label{dhectr}
\eeq
where $j_{max}$ and $j'_{max}$ depend on the specific family, 
the $A_j$ and $b_{j'}$ are polynomials in $q$, and their dependence on 
$k_1$ and $k_2$ is left implicit.  (This notation follows
that of Ref. \cite{strip} except that in eq. (\ref{nhectr}), we use $A$ rather
than $a$ to avoid confusion with the $a_j$ functions in eq. (\ref{pgsum}).) 
A general formula for $A_0$ is
\beq
A_0 = P(T_{k_1+k_2-1},q) = q(q-1)^{k_1+k_2-2}
\label{a0hectr}
\eeq
The other $A_j$'s and $b_{j'}$ polynomials will be given below for specific
families. 

For the starting family, $HEC_{0,0}(\overline K_2 + T_r) = \overline K_2 +
T_r = (K_2)_{b=1} + T_r$, and indeed its generalization 
$(K_p)_b + T_r$, one has \cite{wa} 
\beq
P((K_p)_b + T_r,q) = q^{(p+1)}\Bigl [(q-p-1)^{r-1}+b(q-p)^{r-2} \Bigr ]
\label{e2tr}
\eeq
(recall the notation in eq. (\ref{ff})). 
The equivalent representation in terms of a generating function is obtained by
noting that eq. (\ref{e2tr}) has the form of eq. (\ref{pgsum}) with $c_0=0$ 
and $N_a=2$, i.e., 
\beq
P((K_p)_b + T_r,q) = c_1(a_1)^r + c_2(a_2)^r
\label{eptrpgsum}
\eeq
with 
\beq
a_1 = q-p-1 \ , \quad a_2 = q-p
\label{a1eptrpgsum}
\eeq
\beq
c_1 = \frac{q^{(p+1)}}{q-p-1}
\label{c1eptrpgsum}
\eeq
and
\beq
c_2 = \frac{bq^{(p+1)}}{(q-p)^2}
\label{c2eptrpgsum}
\eeq
We find that the chromatic polynomial $P((K_p)_b + T_r,q)$ is given as
the coefficient of $x^r$ in the Taylor series expansion about $x=0$ of the
generating function 
\beq
\Gamma \Big ( (K_p)_b + T_r,q;x \Bigr ) = 
\frac{x(A_0 + A_1x)}{(1-\lambda_1x)(1-\lambda_2x)}
\label{gammaeptr}
\eeq
where
\beq
\lambda_j = a_j \ , \quad j=1,2
\label{lambdajeq}
\eeq
\beq
A_0 = a_1c_1 + a_2c_2
\label{a0eq}
\eeq
and 
\beq
A_1 = -a_1a_2(c_1+c_2)
\label{a1eq}
\eeq
For a general graph $G$, this type of relation between the form for the
chromatic polynomial (\ref{eptrpgsum}) and the generating function 
(\ref{gammaeptr}) has a straightforward generalization to the case where 
$N_a \ge 3$.  In the present case $(K_2)_{b=1} + T_r = 
HEC_{0,0}(\overline K_2 + T_r)$, we find, for the upper limits on the sums in 
${\cal N}(x)$ and ${\cal D}(x)$ the values $j_{max}=1$ and $j'_{max}=2$.  

\subsection{Families $HEC_{\lowercase{k}_1-2,0}
(\overline K_{\lowercase{2}} + T_{\lowercase{r}})$}

We first consider the simplest homeomorphic expansion of type $HEC$ on the 
starting family $\overline K_2 + T_r$, in which one performs this expansion
only on the bonds connecting one of the two vertices in the $\overline K_2$
subgraph with the vertices of the $T_r$ subgraph.  With no loss of generality,
we choose this vertex in $\overline K_2$ to be $v_1$, so that $k_2-2=0$ and 
$k_1-2 \ge 1$.  For brevity of notation, define 
\beq
O_{k,r} = HEC_{k_1-2,0}(\overline K_2 + T_r) \ , \quad k_1 \equiv k
\label{okr}
\eeq
with $v(O_{k,r})=2+r(k-1)$.
When referring to the collection of the graphs of this family for various $r$,
we denote it as $\{O_k \}$.  The case $k=2$ is just the original starting
family, $\overline K_2 + T_r$. 
For the family $O_{k,r}$ with $k \ge 3$ we find that 
\beq
j_{max}=2 \ , \qquad j'_{max}=3
\eeq
For general $k \ge 3$, the 
denominator of the generating function can be written as
\beqs
{\cal D}(\{O_k\},q,x) & = & \Bigl [1-(q-2)D_k x\Bigr]
\Bigl [1-(q-3)D_k x -(q-1)(q-2)D_{k-1}D_k x^2 \Bigr ] \cr\cr
& = & (1-\lambda_2 x)(1-\lambda_{1p}x)(1-\lambda_{1m}x)
\label{dgokr}
\eeqs
where
\beq
\lambda_{1p,m} = \frac{1}{2}\Biggl [ (q-3)D_k \pm \Bigl [ \{(q-3)D_k\}^2 
+ 4(q-1)(q-2)D_{k-1}D_k \Bigr ]^{1/2} \Biggr ]
\label{lambda1ok}
\eeq
and
\beq
\lambda_2 = (q-2)D_k
\label{lambda2ok}
\eeq
The coefficient functions in ${\cal N}$ are $A_0$, given by the special case of
eq. (\ref{a0hectr}), viz., $A_0=q(q-1)^k$, and 
\beq
A_1=-q(q-1)D_k\Biggl [(2q-5)(q-1)^{k-1}-(q-2)\Bigl [qD_k-2(-1)^k \Bigr]\Biggr ]
\label{a1ok}
\eeq
\beq
A_2=-q^2(q-1)^2(q-2)D_{k-1}D_k^2
\label{a2ok}
\eeq
(In deriving the expression for $A_1$, we have used the identity
$D_{2k-1}=[qD_k-2(-1)^k]D_k$.)  We calculate 
\beq
W(\{O_k\},q) = (\lambda_{max})^{1/(k-1)} \quad {\rm for} \quad q \in R_1
\label{wokr1}
\eeq
and
\beq
|W(\{O_k\},q)| = |\lambda_{max}|^{1/(k-1)} \quad {\rm for} \quad q \in R_j 
\neq R_1
\label{wokrj}
\eeq
where $\lambda_{max}(q)$ denotes the leading $\lambda$ in eq. (\ref{dgokr})
in the respective regions of the $q$ plane. 
For a general family $\{G\}$ of graphs with a generating function 
\beq
\Gamma(\{G\},q,x) = \frac{{\cal N}(x)}{{\cal D}(x)} = 
\sum_{j=j_0}^{\infty} P(G_r,q)x^r
\label{gammaform}
\eeq
with 
\beq
{\cal D}(x) = \prod_{j'=1}^{j'_{max}}(1-\lambda_j x)
\label{lambdaform}
\eeq
the boundary ${\cal B}$ in the limit $r \to \infty$ is given (see eq. (4.16) of
Ref. \cite{strip}) as solution locus of the 
equation expressing the degeneracy in magnitude of two leading 
$\lambda$ terms in eq.\ (\ref{lambdaform}), where $W$ switches form
nonanalytically,
\beq
| \lambda_{max}(q)| = | \lambda_{max'}(q)|
\label{degeneracylam}
\eeq

As an illustration, we consider the lowest homeomorphic expansion, $k=3$, for
which eqs. (\ref{lambda1ok}) and (\ref{lambda2ok}) read 
\beq
\lambda_{1p,m} = \frac{1}{2}(q-2)\biggl [ (q-3) \pm 
\Bigl (q^2-2q+5 \Bigr )^{1/2} \biggr ]
\label{lambda1okrkeq3}
\eeq
\beq
\lambda_2 = (q-2)^2
\label{lambda2okrkeq3}
\eeq
Let us now write $\lambda_{1p}$, $\lambda_{1m}$ and $\lambda_2$ in terms 
of the variable $z=1/q$ in polar coordinates, as given by eq. (\ref{zpolar})
and Taylor-expand the resulting expressions for small $\zeta$. We obtain for
the magnitudes squared of the $\lambda$'s
\beq
|\lambda_{1p}|^2 = 1-4\zeta \cos\theta+2\zeta^2(\cos(2\theta)+2)+O(\zeta^3)
\label{lam1pok3polar}
\eeq
\beq
|\lambda_{1m}|^2 = \zeta^2 + O(\zeta^3)
\label{lam1mok3polar}
\eeq
\beq
|\lambda_2|^2 = 1-4\zeta \cos\theta +4\zeta^2 + O(\zeta^3)
\label{lam2ok3polar}
\eeq
Thus, for small values of $|z|$, 
the boundary ${\cal B}$ is given by the equation
$|\lambda_{1p}|=|\lambda_2|$, which yields $\zeta^2\cos(2\theta)+O(\zeta^3) =
0$ as $\zeta \to 0$.  
Hence, ${\cal B}$ is noncompact in the $q$ plane and crosses the 
origin of the $z$ plane at angles such that $\cos(2\theta)=0$, i.e., 
\beq
\theta= \frac{\pi}{4} + n \frac{\pi}{2} \ , \qquad 0 \le n \le 3 
\label{anglesokeq3}
\eeq
This is evident in Fig. \ref{heck32tr}. From eq. ({\ref{y}), it follows that 
$y \to z$ as $z \to 0$, so that these angles are the same in the $y$ plane. 
The boundary ${\cal B}$ divides the $z$ plane into four regions: 
(i) $R_1$, including the interval $0 < z < 1/3$ 
on the real $z$ axis; (ii) $R_2$, including the intervals $-\infty < z < 0$ and
$1/3 < z < \infty$ on the real $z$ axis; and (iii) the complex-conjugate pair
of regions $R_3$ and $R_3^*$ lying roughly above and below $z=0$.   

We find
\beq
W([\lim_{r \to \infty} O_{3,r}],q) = 
(\lambda_{1p})^{1/2} \quad {\rm for} \quad 
q \in R_1
\label{wo3r1}
\eeq
\beq
|W([\lim_{r \to \infty} O_{3,r}],q)| = |\lambda_{1p}|^{1/2} \quad {\rm for} 
\quad q \in R_2
\label{wo3r2}
\eeq
and
\beq
|W([\lim_{r \to \infty} O_{3,r}],q)| = |\lambda_2|^{1/2} \quad {\rm for} \quad 
q \in R_3 \ , R_3^*
\label{wo3r1r3}
\eeq
A portion of ${\cal B}$ crosses the positive real axis at $z=z_c=1/3$ so that 
$q_c = 3$ for $\lim_{r \to \infty} O_{3,r}$.  Along this portion of the
boundary $|\lambda_{1m}|$ becomes degenerate with $|\lambda_{1p}|$, although
the former never dominates in magnitude over the latter.  This portion of the
boundary ends in two c.c. multiple points.  Note that
\beq
W([\lim_{r \to \infty} O_{3,r}],q) = 0 \quad {\rm at} \quad q=2
\label{wo3q2}
\eeq
consistent with eqs. (\ref{chiheceptr}) and (\ref{peptrq20}) 

\begin{figure}
\centering
\leavevmode
\epsfxsize=3.5in
\begin{center}
\leavevmode
\epsffile{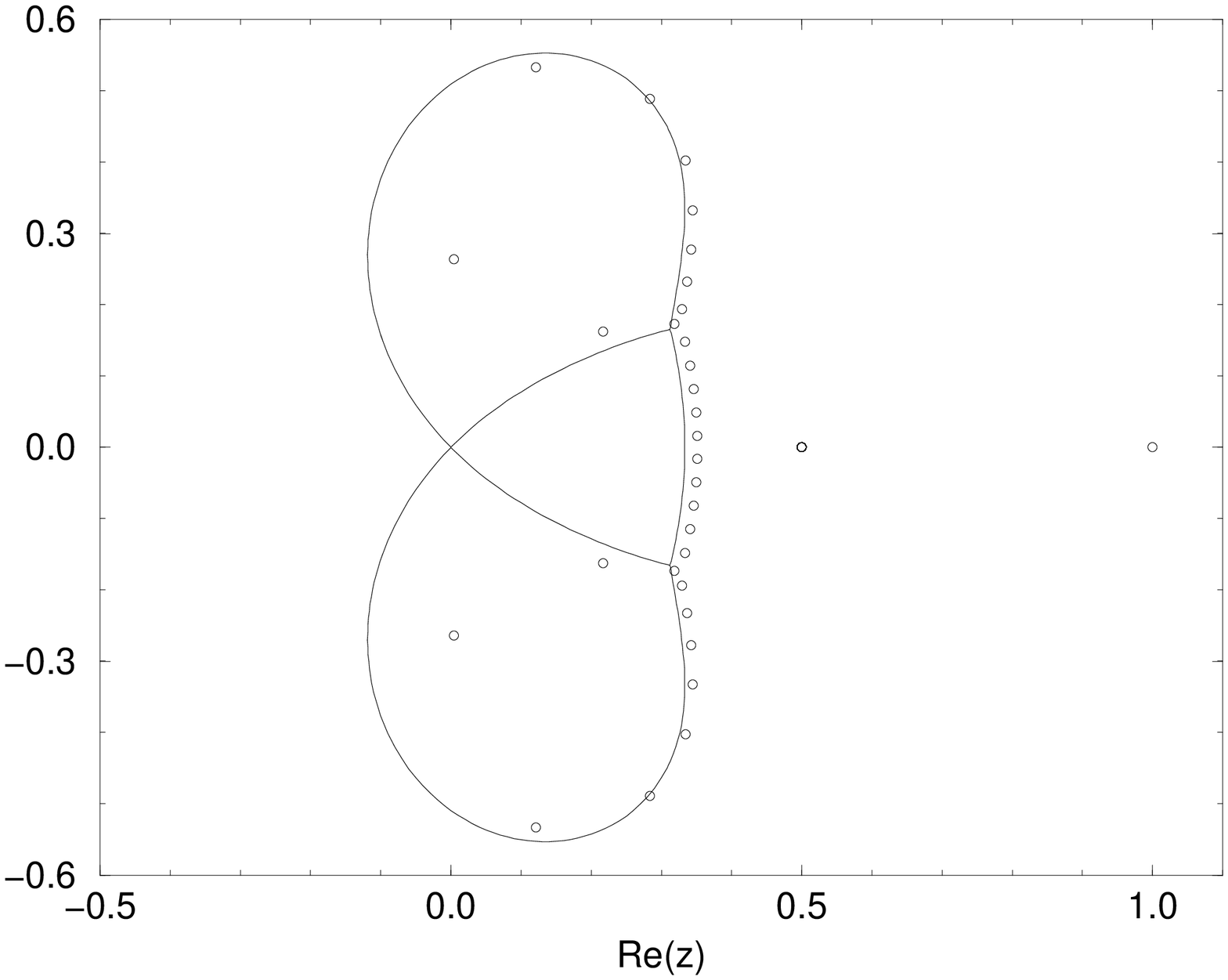}
\end{center}
\vspace{-2cm}
\caption{\footnotesize{Boundary ${\cal B}$ in the $z=1/q$ plane for 
$\lim_{r \to \infty} O_{k,r}$ with $k=3$. Chromatic zeros for $O_{k,r}$ 
$(k,r)=(3,29)$ are shown for comparison.}}
\label{heck32tr}
\end{figure}

\subsection{Families $HEC_{\lowercase{k}_1-2,\lowercase{k}_2-2}
(\overline K_{\lowercase{2}} + T_{\lowercase{r}})$ with $k_1=k_2$}

We next consider the case of symmetric homeomorphic expansion of
the bonds from the two vertices of the $\overline K_2$ subgraph, i.e., 
$k_1=k_2 \equiv k$.  For brevity of notation, we denote 
\beq
U_{k,r} = HEC_{k_1-2,k_2-2}(\overline K_2 + T_r) \ , \quad k_1 = k_2 \equiv k
\label{ukr}
\eeq
with $v(U_{k,r})=2+r(2k-3)$. 
When referring to the collection of the graphs of this family for various $r$,
we denote it as $\{U_k \}$.

For the family $U_{k,r}$ we find that 
\beq
j_{max}=3 \ , \quad j'_{max}=4
\label{jjp}
\eeq
The denominator of the generating function can be written as
\beqs
{\cal D}(\{U_k\},q,x) & = & 
\Bigl [ 1 - (q-2)D_k^2 x - (q-1)^3 D_{k-1}^2 D_k^2 x^2 \Bigr ] \times \cr\cr
& & \Bigl [ 1-D_k(D_{k+1}-D_k)x-(q-1)^2D_{k-1}D_k^3x^2 \Bigr ]
\label{duk}
\eeqs
Observe that each factor of $x$ in ${\cal D}(\{U_k\},q,x)$ occurs with at least
one accompanying factor of $D_k$.  We have 
\beq
{\cal D}(\{U_k\},q,x) = (1-\lambda_{1p}x)(1-\lambda_{1m}x)(1-\lambda_{2p}x)
(1-\lambda_{2m}x)
\label{duklambda}
\eeq
where
\beq
\lambda_{1p,m} = \frac{D_k}{2}\Biggl [ (q-2)D_k \pm \Bigl [ [(q-2)D_k]^2
+ 4(q-1)^3(D_{k-1})^2 \Bigr ]^{1/2} \ \Biggr ]
\label{lam1pmuk}
\eeq
and
\beq
\lambda_{2p,m} = \frac{D_k}{2} \Biggl [ (D_{k+1}-D_k) \pm \Bigl [ 
(D_{k+1}-D_k)^2 + 4(q-1)^2D_{k-1}D_k \Bigr ]^{1/2} \ \Biggr ]
\label{lam2pmuk}
\eeq
For the polynomials in the numerator ${\cal N}(x)$, eq. (\ref{a0hectr}) gives 
$A_0 = q(q-1)^{2k-2}$ and we find 
\beq
A_1 = q(q-1)\Biggl [(D_{2k-1})^2 -(q-1)^{2k-3}\Bigl [(q-3)D_k+D_{k+1} 
\Bigr ]D_k \Biggr ]
\label{a1uk}
\eeq
\beqs
A_2 &=& -q(q-1)^2D_{k-1}D_k^3 \Bigl [q^2D_k^2-q(q+2)(-1)^kD_k +2q+\cr\cr
& & q(q-2)(q-1)^{k-2}D_k+(q-2)(-1)^k(q-1)^{k-2}+1 \Bigr ]
\label{a2uk}
\eeqs
\beq
A_3 = -q^3(q-1)^4D_{k-1}^3D_k^5
\label{a3uk}
\eeq

We calculate 
\beq
W(\{U_k \},q) = [\lambda_{max}(q)]^{1/(2k-3)} \quad {\rm for} \quad 
q \in R_1
\label{wlambdaukr1}
\eeq
and 
\beq
|W(\{U_k \},q)| = |\lambda_{max}(q)|^{1/(2k-3)} \quad {\rm for} \quad 
q \in R_j \neq R_1
\label{wlambdaukotherr}
\eeq
where $\lambda_{max}(q)$ refers to the respective dominant $\lambda$ in the
given region.  The boundary ${\cal B}$ is given, as before, by the solution 
locus of eq. (\ref{degeneracylam}).  We show this boundary in Fig. 
\ref{hectr}(a), together with illustrative chromatic zeros for the case $r=19$.
 From the property (\ref{dkkoddfactor}), 
it follows that the $\lambda$'s are degenerate at zero when $q=2$, 
\beq
\lambda_{1p,m}(q=2) = \lambda_{2p,m}(q=2)=0 \quad {\rm for} \quad k \quad 
{\rm odd}
\label{lambdakoddq2}
\eeq
so that ${\cal B}$ passes through the point $q=2$ if $k$ is odd, and, 
furthermore, 
\beq
W([\lim_{r \to \infty}U_{k,r}],q=2) = 0 \quad {\rm for} \quad k \quad {\rm odd}
\label{wkoddq20}
\eeq
This is consistent with the facts that $\chi(\{U_k\})=3$ and 
$P(U_{k,r},q=2)=0$ as special cases of eqs. (\ref{chiheceptr}) and 
(\ref{peptrq20}). 

We next give some explicit results for the lowest two cases, $k=3$ and $k=4$.  
For the family $U_{k=3,r}$ eqs. (\ref{lam1pmuk}) and (\ref{lam2pmuk}) yield 
\beq
\lambda_{1p,m;k=3} = \frac{1}{2}(q-2)\biggl [ (q-2)^2 \pm 
\Bigl (q^4-4q^3+12q^2-20q+12 \Bigr )^{1/2} \biggr ]
\label{lambda1ukrkeq3}
\eeq
\beq
\lambda_{2p,m;k=3} = \frac{1}{2}(q-2)\biggl [ (q^2-4q+5) \pm 
\Bigl (q^4-4q^3+10q^2-20q+17 \Bigr )^{1/2} \biggr ]
\label{lambda2ukrkeq3}
\eeq
To investigate the boundary in the vicinity of $z=0$, we divide the degeneracy
equations by $|q|^3$ and express the results in terms of $z=1/q$ in polar 
coordinates, as given by eq. (\ref{zpolar}).  We then 
Taylor-expand these equations for small $\zeta=|z|$. This yields 
\beq
|\lambda_{1p;k=3}|^2 = 4-24\zeta \cos\theta + 4\zeta^2 [8\cos(2\theta)+9]
-8\zeta^3 [12 \cos\theta +\cos(3\theta)]+O(\zeta^4)
\label{lam1puk3polar}
\eeq
\beq
|\lambda_{2p;k=3}|^2 = 4-24\zeta \cos\theta + 4\zeta^2 [8\cos(2\theta)+9]
-16\zeta^3 [6 \cos\theta +\cos(3\theta)]+O(\zeta^4)
\label{lam2puk3polar}
\eeq
\beq
|\lambda_{1m;k=3}|^2 = 4\zeta^2 + O(\zeta^3)
\label{lam1muk3polar}
\eeq
\beq
|\lambda_{2m;k=3}|^2 = 4\zeta^2 + O(\zeta^3)
\label{lam2muk3polar}
\eeq
Thus, in the vicinity of $z = 0$ the boundary ${\cal B}$ is given by the 
equation $|\lambda_{1p;k=3}|=|\lambda_{2p;k=3}|$, which yields 
$\cos(3\theta) = 0$ 
for $\zeta\neq 0$.  Hence, six curves on ${\cal B}$ (forming three branches) 
cross the point $z=0$ (and hence also the point $y=0$), at angles
\beq
\theta= \frac{\pi}{6} + n \frac{\pi}{3} \ , \qquad 0 \le n \le 5
\label{anglesukeq3}
\eeq
The boundary ${\cal B}$ divides the $z$ plane into six regions.  As one
traverses a circle around the origin, $z=0$, starting with a small positive
real $z$ value, first $\lambda_{1p;k=3}$ is dominant, and then the dominant 
$\lambda$'s alternate between $\lambda_{1p;k=3}$ and $\lambda_{2p;k=3}$. 
The resultant $W$ functions are
given by eqs. (\ref{wlambdaukr1}) and (\ref{wlambdaukotherr}). 

For the family $U_{k=4,r}$ eqs. (\ref{lam1pmuk}) and (\ref{lam2pmuk}) yield
\beq
\lambda_{1p,m;k=4} = \frac{1}{2}(q-2)(q^2-3q+3)\biggl [ (q^2-3q+3) \pm 
\Bigl (q^4-2q^3+3q^2-6q+5 \Bigr )^{1/2} \biggr ]
\label{lambda1ukrkeq4}
\eeq
\beqs
\lambda_{2p,m;k=4} & = & \frac{1}{2}(q^2-3q+3)\biggl [ (q^3-5q^2+9q-7) \pm
\cr\cr
& & 
\Bigl (q^6-6q^5+15q^4-24q^3+35q^2-42q+25 \Bigr )^{1/2} \biggr ]
\label{lambda2ukrkeq4}
\eeqs
As before, the boundary ${\cal B}$ is determined by eq. (\ref{degeneracylam}).
This boundary is plotted in Fig. \ref{hectr}(b). 
Let us write $\lambda_{1p,m}$ and $\lambda_{2p,m}$ in terms 
of the variable $z=1/q$ in polar coordinates, as given by eq. (\ref{zpolar})
and Taylor expand the resulting expressions for small $\zeta$. We find that the
two $\lambda$'s that are dominant near $z=0$, namely $\lambda_{1p}$ and 
$\lambda_{2p}$, have squared magnitudes that coincide through $O(\zeta^4)$ and
differ in order $O(\zeta^5)$ in the coefficient of $\zeta^5 \cos(5\theta)$.  
(The other two 
$\lambda$'s are subdominant near $z=0$; indeed, they vanish there, as 
$|\lambda_{1m}|^2 \sim |\lambda_{2m}|^2 = 4\zeta^2 + O(\zeta^3)$.)
It follows that as $\zeta \to 0$, 
the boundary ${\cal B}$ is given by the equation
$|\lambda_{1p}|=|\lambda_{2p}|$, which yields $\cos(5\theta) = 0$ for 
$\zeta\neq 0$. Hence, ten curves, forming five branches of ${\cal B}$, cross
the point $z=0$ (and also $y=0$) at the angles 
\beq
\theta= \frac{\pi}{10} + n \frac{\pi}{5} \ , \qquad 0 \le n \le 9
\label{anglesukeq4}
\eeq
Here, the boundary ${\cal B}$ divides the $z$ plane into eight regions. As one
traverses a circle around the origin starting with a small positive real value
of $z$, first $\lambda_{1p;k=4}$ is dominant, and then the dominant 
$\lambda$'s alternate between $\lambda_{1p;k=4}$ and $\lambda_{2p;k=4}$, as
before with $k=3$.  The resultant $W$ functions again follow from eqs. 
(\ref{wlambdaukr1}) and (\ref{wlambdaukotherr}).  
Thus, comparing our results for 
$U_{k,r}$ with $k=3$ and $k=4$, we observe that the number of curves on 
${\cal B}$ passing through $z=0$, and the number of regions, increase as $k$ is
increased.  One should also remark that certain properties of the region
depend on whether $k$ is even or odd, such as the fact that ${\cal B}$ passes
through $z=1/2$ for odd $k \ge 3$.  

\pagebreak
\begin{figure}
\centering
\leavevmode
\epsfxsize=3.5in
\begin{center}
\leavevmode
\epsffile{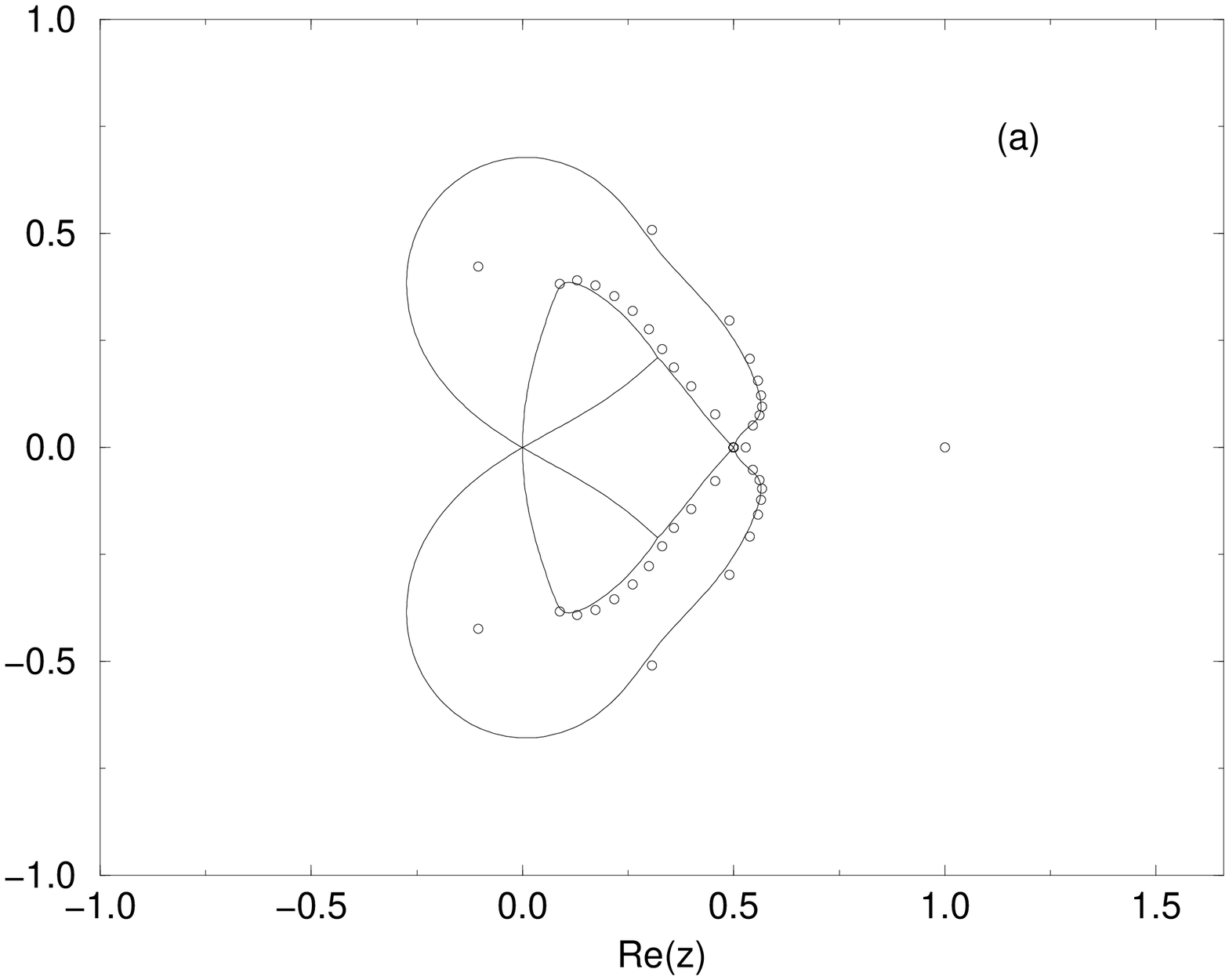}
\end{center}
\vspace{-3cm}
\begin{center}
\leavevmode
\epsfxsize=3.5in
\epsffile{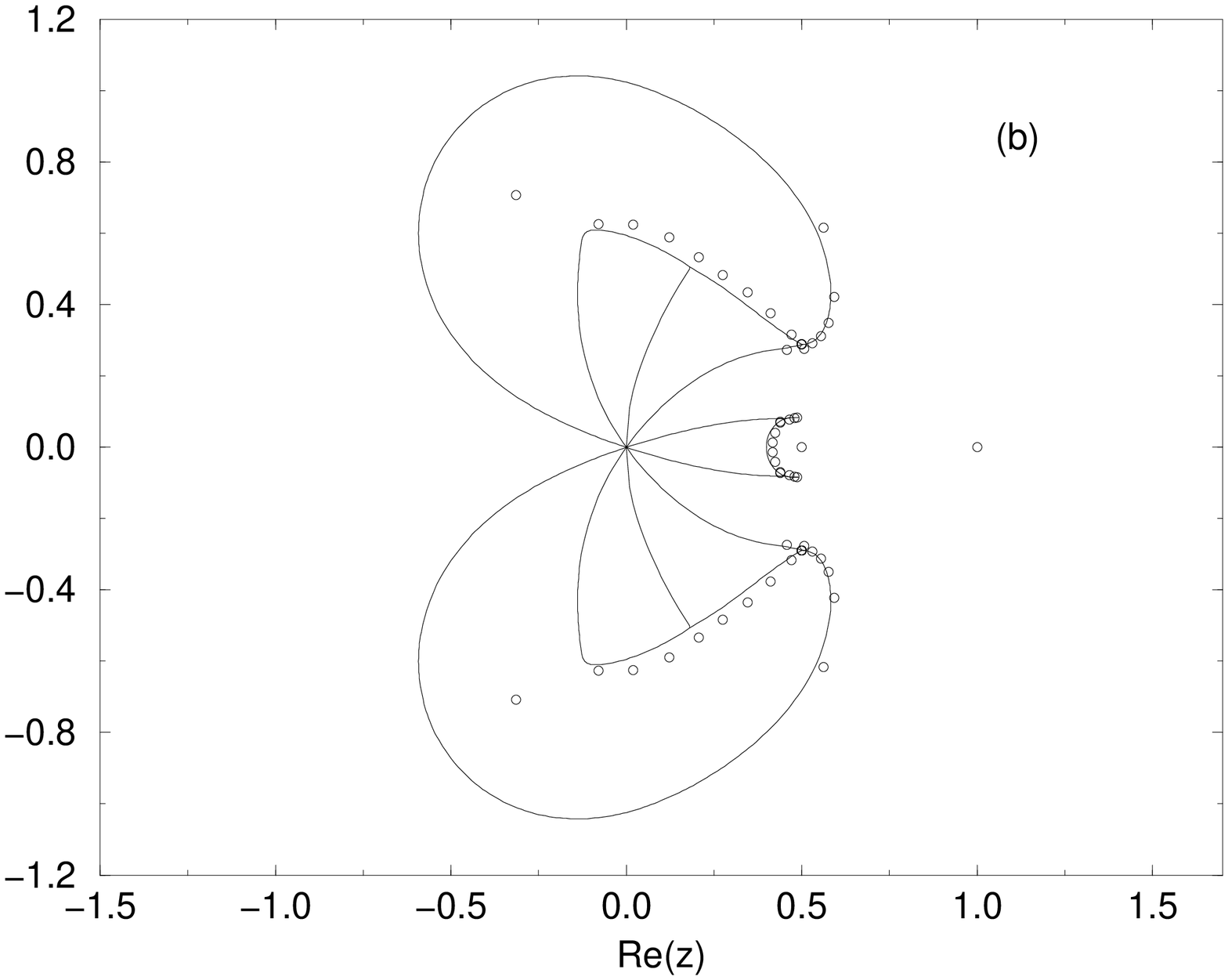}
\end{center}
\vspace{-2cm}
\caption{\footnotesize{Boundary ${\cal B}$ in the $z=1/q$ plane for 
$\lim_{r \to \infty}U_{k,r}$ with $k=$ (a) 3 (b) 4.  
Chromatic zeros for $U_{k,r}$ with $(k,r)=$ (a) (3,19) (b) (4,15) are shown 
for comparison.}}
\label{hectr}
\end{figure}

\section{Discussion}

In this section we discuss general features of the Potts antiferromagnet $W$
functions on all of the families of graphs that we have constructed and studied
in this paper and our earlier work, with loci ${\cal B}$ that are noncompact in
the $q$ plane.  In (the theorem of section IV of)
Ref. \cite{wa} we gave the general algebraic condition that for a particular
family of graphs, the infinite--vertex limit yields a locus ${\cal B}$ that
passes through $z=0$.  From our calculations we have observed
a geometrical regularity in the families of graphs that have this
property, viz., that in this infinite-vertex limit they all contain an infinite
number of different, non-overlapping (and non-self-intersecting) circuits, each
of which passes through at least two fixed, nonadjacent vertices. This
immediately implies that these aforementioned nonadjacent vertices have degrees
$\Delta$ that go to infinity in this limit.  One thus sees at the graphical
level why the derivation of the large--$q$ series for the reduced function
$W_{red}(\{G\},q)$ or equivalently $\overline W(\{G\},q)$ on
regular lattices with free or some type of periodic boundary conditions works;
in the thermodynamic limit, such lattices do not have the two or more
nonadjacent vertices with degrees $\Delta \to \infty$ as in the criterion
stated above for noncompact ${\cal B}$.

We remark that in studies of the 
Potts model on regular lattices, it has been useful to utilize certain boundary
conditions that do involve vertices which, in the thermodynamic limit, have
infinite degree $\Delta$, 
since these make it possible to preserve exact duality
on finite lattices \cite{martin,wuprl,pfef}.
On the square lattice, the duality-preserving boundary 
conditions (DBC's) of types DBC-1 and DBC-2, in the notation of Ref. 
\cite{pfef} involve one such vertex with $\Delta \to \infty$ while the DBC-3
type involves two such vertices; however, in the latter case, these are
adjacent.  Hence, none of these duality-preserving boundary conditions
invalidates a large--$q$ expansion of $W_{red}(\{G\},q)$ for this lattice.

An elementary topological property should be noted: for an arbitrary family of
graphs, the continuous locus of points ${\cal B}$ where $W$ is nonanalytic in
the infinite--vertex limit does not necessarily enclose regions in the $q$
plane.  Indeed, in Ref. \cite{strip} we carried out exact 
calculations of a variety of infinitely long, finite-width strips of different
lattices and found loci ${\cal B}$ that consisted of arcs (and line segments)
that did not enclose regions.  (We have also performed similar calculations for
other types of strips with specific types of end-graphs that yield loci ${\cal
B}$ that do enclose regions.)  None of these loci were noncompact in the $q$
plane.  In contrast, if the infinite-vertex limit of a family of graphs yields
a locus ${\cal B}$ that passes through $z=0$, then, given that it is not a
semi-infinite line segment on the positive or negative $z$ plane, it 
necessarily separates the $z$ and equivalently the $q$ or $y$ planes into at
least two different regions where $W$ is an analytic function.  In
Ref. \cite{strip}, we observed that the endpoints of the arcs (or line segments
on the real $q$ axis) that comprised the nonanalytic locus ${\cal B}$ were
determined by the branch points of certain algebraic expressions occurring in
the $\lambda$'s that entered in degeneracy equations of the form 
of eq. (\ref{degeneracylam}).  Since, for example, the $\lambda$'s for the
family $\{U_k\}$, eqs. (\ref{lam1pmuk}) and (\ref{lam2pmuk}), also contain 
square roots, one may wonder what role the branch points of these roots play
with regard to the boundary ${\cal B}$.  We have investigated this and have
found that the analogues of the arcs extending between branch points of the 
square roots, which comprised the various boundaries in Ref. \cite{strip}, do
not yield endpoint singularities on ${\cal B}$ here.  The reason for this is
that either (i) these arcs involve degeneracy of $\lambda$'s that are
nonleading in a given region, or (ii) although they coincide on part of their
length with the actual boundary, the portion of the arcs containing the
endpoints lies off this boundary, in a region where condition (i) holds.  This
is illustrated in Fig. \ref{arcsuk3} for the $\{U_k\}$ family with $k=3$, where
we show the actual boundary ${\cal B}$, as in Fig. \ref{hectr}(a), together
with the arcs (drawn in a thicker black) formed by the degeneracy conditions 
\beq
|\lambda_{2p}|=|\lambda_{2m}|
\label{lambda2p2muk3}
\eeq
and 
\beq
|\lambda_{1p}|=|\lambda_{1m}|
\label{lambda1p1muk3}
\eeq

\begin{figure}
\centering
\leavevmode
\epsfxsize=3.5in
\begin{center}
\leavevmode
\epsffile{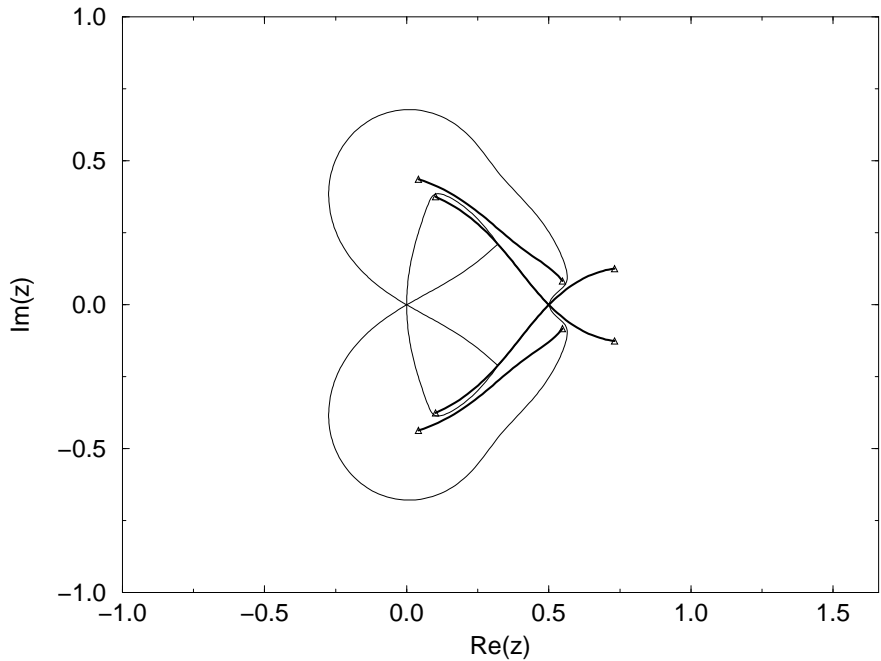}
\end{center}
\vspace{-2cm}
\caption{\footnotesize{Boundary ${\cal B}$ in the $z=1/q$ plane for 
$\lim_{r \to \infty}U_{3,r}$, shown together with the locus of solutions of the
degeneracy equations (\ref{lambda2p2muk3}) and (\ref{lambda1p1muk3}).  See text
for discussion.}}
\label{arcsuk3}
\end{figure}

\noindent

The complex-conjugate pair of arcs which constitute the solution locus to eq. 
(\ref{lambda2p2muk3}), with endpoints at $z_{2a}, \ z_{2a}^* = 
0.04106 \pm 0.43627i$ and $z_{2b}, \ z_{2b}^* =0.54717 \pm 0.08333i$, 
lie in the interior of the regions $R_4$ and $R_4^*$.  But in these regions,
neither $\lambda_{2p}$ nor $\lambda_{2m}$ is dominant; rather, as we discussed
in the section on the $U_{k,r}$ family, it is $\lambda_{1p}$ that is dominant
in these regions.  Hence, this complex-conjugate pair of arcs is not relevant
to the actual boundary. As shown in Fig. \ref{arcsuk3}, the locus of solutions 
for the degeneracy equation (\ref{lambda1p1muk3}) forms a complex-conjugate 
pair of arcs that cross each other and the real $z$ axis at $z=1/2$ and have
endpoints at $z_{1a}, \ z_{1a}^* = 0.10169 \pm 0.37529i$ and 
$z_{1b}, \ z_{1b}^* = 0.73164 \pm 0.12612i$. The portion of these arcs that
lie to the right of $Re(z) = 1/2$ are not relevant for determining the boundary
because in this region, denoted $R_2$ in our section above on the $\{U_k\}$
family, the dominant $\lambda$ is $\lambda_{2p}$.  The portion of the
complex-conjugate arcs lying between the multiple (crossing) point $z=1/2$ on 
the right and the multiple points forming T-intersections on the left at 
$z_T, \ z_T^*=0.3204 \pm 0.2110i$ 
does coincide with the boundary, since on this portion the boundary is
determined by the degeneracy of magnitudes of leading eigenvalues 
$\lambda_{1p}$ and $\lambda_{1m}$, i.e., eq. (\ref{lambda1p1muk3}). However, to
the left of these T-intersection points, the arcs forming the solution locus of
eq. (\ref{lambda1p1muk3}) again deviate from the actual boundary, which is
determined by the degeneracy of leading magnitudes 
$|\lambda_{2p}|=|\lambda_{1p}|$.  Hence, once again, the left endpoints of
these arcs are not relevant for ${\cal B}$.   Thus, the boundaries ${\cal B}$
for the families considered here, even aside from the fact that they
automatically enclose regions owing to their noncompactness, in contrast to the
arcs comprising the loci ${\cal B}$ for the strip graphs that we studied in
Ref. \cite{strip}, also differ qualitatively in that they do not contain any
arc endpoints.  We have explained why this is so even though some of the
families do have $\lambda$'s containing branch points singularities.  
This emphasizes that 
the existence of the arc-like structure of the loci ${\cal B}$ that we found in
Ref. \cite{strip} depended 
not just on the presence of branch point singularities in the $\lambda$'s 
entering via equations analogous to eq. (\ref{duklambda}) in the denominators 
${\cal D}$ of the generating functions for various strip graphs, but also on 
the fact that these branch point singularities occurred at endpoints of arcs in
regions where these arcs were the solution loci of magnitude degeneracy
equations for leading $\lambda$'s.  

 From our
studies of many homeomorphic expansions, we have found the general feature that
for a given homeomorphic class parametrized by some set of homeomorphic
expansion indices $\{k_j \}$, the number of regions separated by the locus
${\cal B}$ in the $r \to \infty$ limit is a nondecreasing function of the above
expansion indices.  This is in accord with 
the constraints from algebraic geometry, in particular, the Harnack theorem
\cite{alg}.  The application of this theorem is simplest in the case where
${\cal B}$ is a nonsingular algebraic curve (so that the number of regions
$N_{reg.} = N_{comp.}+1$ where $N_{comp.}$ denotes the number of connected
components of ${\cal B}$), as is the case for the $r \to \infty$ 
limit of the families $T_{k,r}=HEG_{k-2}(\overline K_2 + T_r)$ and 
$S_{k,r}=HEG_{k-2}(\overline K_3 + T_r)$.  In these cases 
(where ${\cal B}$ has no multiple points) the Harnack theorem states that the
number of regions is bounded above by $g+1$, where $g$ is the genus of the
algebraic equation in the variables $Re(q)$ and $Im(q)$ whose solution set is 
${\cal B}$, viz., $g=(d-1)(d-2)/2$ where $d$ is the homogeneous degree of 
this equation.  Since, as our exact solutions show, this degree and the
resultant genus increase as a function of the homeomorphic expansion
indices, the Harnack upper bound also increases.  It should be noted,
however, that the number of regions may remain the same as one performs a
homeomorphic expansion (e.g., $N_{reg.}=2$ for the $r \to \infty$ limit of 
$HEG_{k-2}(\overline K_2 + T_r)$ for $k=2,3$, and 4, then $N_{reg.}=3$ for 
$k=5,6$, and 7, etc.)  

For algebraic curves ${\cal B}$ with
singularities such as multiple points, 
one no longer has the simple relation $N_{reg.} = N_{comp.}+1$.
Indeed, in the families constructed and analyzed in this paper, ${\cal B}$
consists of a single connected component, i.e., $N_{comp.}=1$, while the number
of regions $N_{reg.}$ and the index of the multiple point at $z=0$ increase 
as a function of the indices $\{k_j \}$ describing the homeomorphic 
expansions.  The origin of this increase is the increase in the degree of the
polynomials in $z$ that occur in the degeneracy equations 
relevant for the boundary ${\cal B}$ in the neighborhood of $z=0$ (e.g., eqs. 
(\ref{degeneqy}) and the resultant (\ref{angles}) for $H_{k,r}$, etc.).  

The boundaries ${\cal B}$ found for the families of graphs in the present paper
exhibit some interesting differences with respect to the (similarly 
noncompact) boundaries that we have previously studied \cite{wa}; the latter 
shared three general properties: (i) a single curve on 
${\cal B}$ passes through $z=0$, so that this point is a regular point on
${\cal B}$ as an algebraic curve; (ii) if $q \in {\cal B}$, then 
$Re(q) \ge 0$, i.e. ${\cal B}$ includes support only for $Re(q) \ge 0$, or 
equivalently, for $Re(z) \ge 0$; and (iii) if $z \in {\cal B}$ and $Re(z)=0$, 
then $z=0$.  In contrast, for the families studied in the present paper, the 
homeomorphic expansions, which are all of the type $HEC$, yield, in the 
$r \to \infty$ limit, respective boundaries ${\cal B}$ that differ in
each of these aspects: (i) several branches of the algebraic curve ${\cal B}$ 
pass through the point $z=0$, which is thus a singular (multiple) point on this
curve; (ii) ${\cal B}$ includes support for $Re(q) < 0$ or equivalently $Re(z)
< 0$, and (iii) ${\cal B}$ includes points with $Re(z) = 0$ other than $z=0$
itself.  Some insight into this can be gained by recalling the differences 
in the types of homeomorphic expansions of starting graphs.  Since in the
original starting families such as $(K_p)_b + G_r$ and $\overline K_p + G_r$,
it was the connecting bonds linking the vertices of the $K_p$ subgraph with the
vertices of the $G_r$ subgraph that gave rise to the noncompactness of the
respective boundaries ${\cal B}$, it makes sense that homeomorphic expansions
of these bonds would alter the nature of the boundaries at $z=0$.  Since this
change involves multiple branches of ${\cal B}$ passing through this point at
various angles, as in eqs. (\ref{angles}), (\ref{anglesokeq3}), 
(\ref{anglesukeq3}) and (\ref{anglesukeq4}), 
this also shows why this type of homeomorphic expansion
leads to boundaries that include support for $Re(q) < 0$, but $q \ne 0, \ 
\infty$, i.e., $Re(z) < 0$ but $z \ne \infty, \ 0$.  In contrast, 
since the bonds within the $G_r$ and $K_p$ subgraphs, by themselves, are not 
directly responsible for the noncompactness of ${\cal B}$, it is plausible that
homeomorphic expansions of these bonds would not change the nature of ${\cal
B}$ at $z=0$.  

\section{Boundary for $L_{\lowercase{k}}$ Limit}

Finally, we briefly discuss the boundary ${\cal B}$ that results when one takes
the limit $L_k$ of eq. (\ref{kinf}), i.e., $k \to \infty$ with $r$ and $p$
fixed, for the families studied in this paper.  We consider the nontrivial
range $r \ge 2$ since for $r=1$, families typically degenerate into tree
graphs with ${\cal B} = \emptyset$.  For the families studied here for which we
have obtained the chromatic polynomials for general $k$, including 
$H_{k,r}$, $O_{k,r}$, and $U_{k,r}$, we find that this boundary is simply the 
unit circle $|q-1|=1$.  This is easily understood, since one can
reexpress our formulas for chromatic polynomials in terms of polynomials in
$a=q-1$ using eqs. (\ref{dk}) and (\ref{pck}).  From eq. (\ref{pck}), it
follows that the chromatic polynomials are of the form of eq. (\ref{pgsum})
with $a$ as the quantity entering in terms raised to respective powers 
proportional to $k$, together with possible $c_0$ terms involving $(-1)^k$, 
and hence, as discussed in our earlier work \cite{w}, ${\cal B}$ is 
determined by the condition $|a|=1$.  Hence, $q_c=2$ and ${\cal B}$ has no 
support for $Re(q) < 0$.  Since the locus $|q-1|=1$ is compact in the $q$ plane
and since our focus here is on the situation where ${\cal B}(q)$
is noncompact in the $q$ plane, passing through $z=0$, and the connection with
the existence of large--$q$ series for $W_{red.}$, the $L_k$ limit is not of 
primary interest here.  We mention, however, that ${\cal B}$ 
divides the $q$ plane into two regions,
$(R_1)_{L_k}$ and $(R_2)_{L_k}$, which are the exterior and interior of the 
unit circle $|q-1|=1$ (the subscripts $L_k$ are appended to 
distinguish these regions from those for the $L_r$ limit). 
It is straightforward to calculate the $W$ function in
these two regions in the $L_k$ limit of the various families.  
For example, for the $H_{k,r}$ family (for fixed $r \ge 2$), 
\beq
W([\lim_{k \to \infty}H_{k,r}],q)=q-1 \quad {\rm for} \quad q \in (R_1)_{L_k}
\label{wkinfr1}
\eeq
and
\beq
|W([\lim_{k \to \infty}H_{k,r}],q)|=1 \quad {\rm for} \quad q \in (R_2)_{L_k}
\label{wkinfr2}
\eeq

\section{Conclusions}

In this paper we have presented exact calculations of the zero-temperature 
partition function, 
$Z(G,q,T=0)$, and ground-state degeneracy (per site), $W(\{G\},q)$, for the
$q$-state Potts antiferromagnet on a number of families of graphs $\{G\}$
for which the boundary ${\cal B}$ of regions of analyticity of $W$ in the
complex $q$ plane is noncompact and has the properties that (i) in the
$z=1/q$ plane, the point $z=0$ is a multiple point on ${\cal B}$; (ii)
${\cal B}$ includes support for $Re(z) < 0$; and (iii) ${\cal B}$ crosses the
imaginary $z$ axis away from $z=0$.  Our results yield insight into the 
conditions which preclude the existence of 
large--$q$ Taylor series expansions for the reduced function 
$W_{red.}=q^{-1}W$.  This insight is valuable since large--$q$ expansions,
where they exist, are of great utility in obtaining approximate values of the
exponent of the ground state entropy, $W$.  

\vspace{10mm}

This research was supported in part by the NSF grant PHY-93-09888.

\vfill
\eject

\end{document}